# Solving the thoracic inverse problem in the fruit fly

Arion Pons[1,2], Illy Perl[1,2], Omri Ben-Dov[1,2,3], Roni Maya[1,2], and Tsevi Beatus[1,2,*]

[1] The Benin School of Computer Science and Engineering, The Hebrew University of Jerusalem, Israel
[2] The Silberman Institute of Life Sciences, The Hebrew University of Jerusalem, Israel
[3] Current address: Max Planck Institute for Intelligent Systems, Tübingen, Germany
* tsevi.beatus@mail.huji.ac.il

**Abstract:** In many insect species, the thoracic structure plays a crucial role in enabling flight. In the dipteran indirect flight mechanism, the thorax acts as a transmission link between the flight muscles and the wings, and it is often thought to act as an elastic modulator: improving flight motor efficiency thorough linear or nonlinear resonance. But as peering closely into the drivetrain of tiny insects is experimentally difficult, the nature of this elastic modulation, and any associated resonant effects, are unclear. Here, we present a new inverse-problem methodology to surmount this difficulty. In a data synthesis process, we integrate experimentally-observable aerodynamic and musculoskeletal data for the fruit fly *D. melanogaster*, and use this integrated dataset to identify several surprising properties of the fly's thorax. We find that fruit flies have an energetic need for flight motor resonance: energy savings due to flight motor elasticity can be up be to 30%. However, the elasticity of the flight muscles themselves is sufficient – and sometimes more than sufficient – to ensure flight motor resonance. In fruit flies, the role of the thorax as an elastic modulator is likely to be insignificant. We discover also a fundamental link between the fruit fly wingbeat kinematics and musculature dynamics: wingbeat kinematics show key adaptions, including in wing elevation angle, that ensure that wingbeat load requirements match musculature load output capability. Together, these newly-identified properties lead to novel conceptual model of the fruit fly's flight motor: a strongly-nonlinear structure, resonant due to muscular elasticity, and intensely concerned with ensuring that the primary flight muscles are operating efficiently. Our inverse-problem methodology sheds new light on the complex behavior of these tiny flight motors, and provides avenues for further studies in a range of other insects.

## 1. Introduction

The flight motors of insects are complex structures. Flight muscles, of varying forms, interact with thoracic structures and additional musculature to generate finely-controlled multi-axis wingbeat motion – in a process showing considerable diversity across phylogenetic orders [1,2]. Studying this process is challenging. Flying insects are small, and their wingbeat motion is rapid. Even with state-of-the-art observation techniques, such as micro-CT [3,4], X-ray diffraction [5–7], sophisticated microscopy [8–10] and high-speed videography [11–19], the roles played by many elements within the flight motor are not well understood. Understanding the propagation of mechanical quantities – deformation, forces, torques, power and energy – is also challenging. Such quantities are only measurable at particular locations, or under particular *ex vivo* conditions: thoracic deformation [20,21]; wing aerodynamics [17,22–28]; muscular strain and load [5,9,29–31]. Attempting to understand the mechanical operation of insect flight motors is thus an *inverse problem* – a problem in which the operation of the motor must be inferred from its inputs and outputs.

In about three-quarters of known insect species, the primary flight muscles are asynchronous, that is, stretch-activated [1,32], and the flight motor inverse problem takes a form of particular interest. In many such insects, the flight mechanism is indirect: the thoracic structure is thought to act as a dynamic transmission between the asynchronous flight muscles and the wing, modulating muscular action through elastic effects [4,33–36]. This thoracic elastic modulation could represent a state of structural resonance – in which thoracic elasticity would absorb wing inertial loads, thus saving energy. Current studies have demonstrated broad evidence for the existence of resonant effects in several insect species, including demonstrations of resonant tuning effects in asynchronous muscles *ex vivo* [37], and in insects with wing and thorax mass alteration [21,38,39]; as well as dynamic mechanical analysis indicating that the thoracic resonant frequency in honeybees (*Apis mellifera*) corresponds approximately to the wingbeat frequency [40]. However, there is also counter-evidence that, in hawkmoths (*Manduca sexta*), resonance may not be significant [41]. Additional studies [42–44] indicate that the elastic behaviour of insect flight motors can be quite complex: in practice, resonant states may exist over a cluster of frequencies, depending on aerodynamic damping and elasticity distribution through the motor [43].

Understanding the role of flight motor elasticity, and the details of thoracic load and power transmission, is important: slight variations in the operation of the flight motor can generate significant changes in flight characteristics [4,10,33,45]. With limited information, the treatment of flight motor elasticity in other areas of insect flight analysis – energetics, control and biomimicry – is restricted. In the study of the energetics of flight, thoracic energetic effects must be assumed *e.g.* via stroke-averaged potential model [46,47]; or negative-work storage mechanism [48–50]; or simple linear elasticity [42]. In the study of insect flight control, correlations between control muscle activity and wing motion are available [3,10,51,52], but the role of thoracic modulation cannot yet be isolated. In the design of biomimetic flapping-wing micro-air-vehicles (FW-MAVs), a range of elastic energy-storage systems have demonstrated their utility [53–57]: with more information on the elasticity properties of insect flight motors, these biomimetic flight motors may be refined even further.

It is in this context that the solution of the flight motor inverse problem becomes attractive. Integration of data for kinematics and loads of both the wings and muscles could itself reveal the elastic modulation of the thorax – and so, its energetic effects, and the implications for FW-MAV systems. In this work we perform this integration and develop a solution process for the flight motor inverse problem. This solution process is similar to a data synthesis process, combining aerodynamic, wingbeat kinematic, and muscular data in the literature. It generates not only the capability for high-fidelity predictions of the dynamics and energetics of flight motor systems, but the capability for identifying qualitative and quantitative dynamical properties of the thorax via inverse problem solution. Applying this process to *Drosophila melanogaster*, as a case study in Diptera, we find several surprising properties. We observe muscular elasticity to be dominant over thoracic elasticity, and overall flight motor (thoracic + muscular) elasticity to show strain-hardening nonlinearity. The nature of these nonlinearities is fundamentally related to *D. melanogaster* wingbeat kinematics. The energetic motivation for flight motor elasticity is shown to represent a complex optimisation problem, with multiple conflicting energetic objectives, including the need for matching muscular load generation to nonlinear wingbeat dissipation. This matching process provides a new cohesive framework for understanding the complexities of insect wingbeat kinematics. As a case in point, we show how synthetic wingbeat kinematics used frequently in the literature [22,27,58] behave in a radically different way to experimentally-observed wingbeat kinematics [23–25,59,60] in terms of load matching. These synthetic wingbeat kinematics are fundamentally unsuitable for the flight musculature, due to the location of the wake capture drag loads within the wingbeat cycle; whereas biological wingbeat kinematics are well-suited in this respect. These considerations, in turn, can provide a motivation for features of biological wingbeat kinematics which have not previously been explained – in particular, the wing elevation angle, which we hypothesise has a key role in ensuring musculature-wing load matching, by altering the location of wake capture drag loads. In this way, the inverse-problem approach to thoracic dynamics provides fundamental insight into the structures, mechanisms, and physiological choices present in insect flight motors; and the translation of these structures, mechanisms and choices into FW-MAVs.

## 2. Methods
### 2.1. Formulating the thorax dynamics as an inverse problem

The problem of identifying causal mechanisms in a system (*e.g.*, an insect flight motor), based on its observable behaviour (*e.g.*, the flight of an insect) is an inverse problem. For example, inverse-problem approaches have been previously utilised to identify the torsional elasticity of the wing root hinge in *D. melanogaster*, based on observed wing kinematics [11,61]. In the flight motor of *D. melanogaster* more broadly, key points of observability are the wings, the muscles, and the thoracic structure. Figure 1 shows a schematic of these points of observability. Each has an associated dataset of observable information: details are examined in §2.2, but for the purposes of formulating inverse problems, it is sufficient to state their nature as follows. (**1**) *Wing dataset*: wing kinematics, observed by high-speed cameras, and wing aerodynamic loads, computed via computational or similitude modelling. (**2**) *Muscular dataset*: muscular geometry, strains, neural signals, and strain-load profiles. (**3**) *Thoracic dataset*: morphology, estimated deformation in flight, and overall stiffness. The particular difficulty of observing the transmission of load and power through the flight motor motivates the use of an inverse-problem methodology.

With these datasets in mind, two distinct inverse problems for the flight motor can be discerned. Both are based on the principle that the loads required to generate observed wing motion – loads which can be predicted computationally or experimentally – must be equivalent to the loads generated by the flight motor at the wing root. These loads, in turn, represent the interaction between the flight muscles, and the thoracic structure (its elasticity, damping, and inertia). In flight, the thoracic structure transmits and modulates the loads generated by the flight muscles to produce the loads required for wing motion. A schematic of this principle is shown in Fig. 1. The first inverse problem asks the questions: how does the thoracic modulation alter the muscular load required to generate the observed wingbeat motion, assuming the muscles can provide load at *any* waveform? And, what is an optimal thoracic modulation, with respect to some metric? For example, considering thoracic elasticity – what is the optimal thoracic elasticity to minimise the flight motor power requirements, assuming the flight motor can provide the correct load waveform for this optimal state? Also, given differing metrics of optimality – *e.g.* minimising energy consumption, mean load, or peak load – what is the available variation in these metrics that could be ascribed to flight motor elasticity? Assuming the motor in unrestricted in its load output, makes this inverse problem one-sided, in that it draws on only one point of observability: the wing. With wing kinematic and aerodynamic data, the optimal thoracic modulation (*i.e.*, flight motor elasticity) can be identified in a given optimality metric. This identification by optimality may then be associated with actual insect behaviour, assuming that insect evolution has optimised the same performance metric. This approach is utilised in several energetic studies [33,46,49,62] as the basis for inferring optimal states of elastic energy storage in the flight motor.

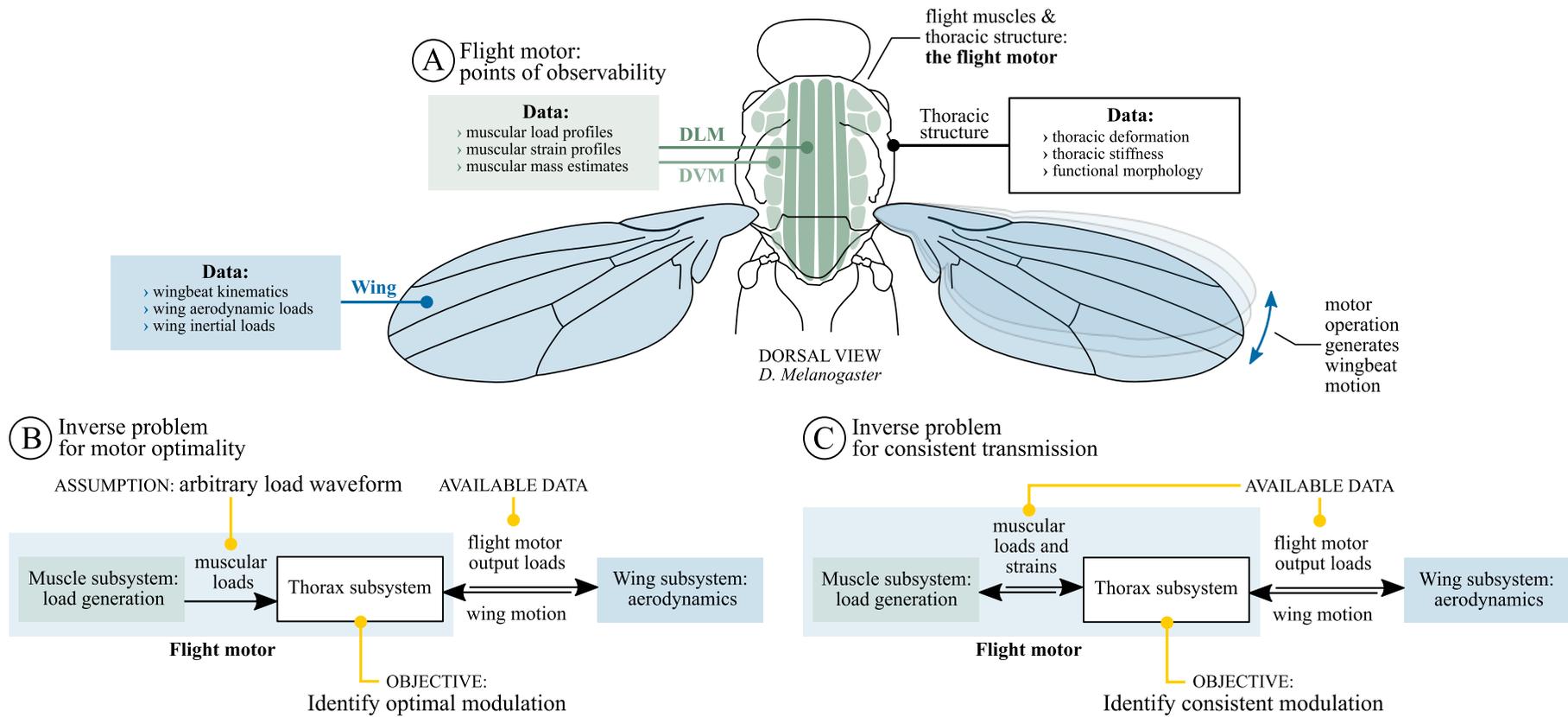

**Figure 1:** Observability and inverse problems in the flight motor of *D. melanogaster*. (**A**) Schematic of the flight motor of *D. melanogaster*, indicating points of observability and associated sources of data. (**B**) Functional block diagram of the inverse problem for optimal thoracic modulation. (**C**) Functional block diagram of the inverse problem for consistent thoracic modulation.

While identification by optimality may seem reasonable, it rests upon assumptions about the optimality of the thoracic behaviour which may or may not be realistic. Hence, the second inverse problem asks the question: given (**i**) the loads generated by the flight muscles; (**ii**) the loads required by the wing; and (**iii**) the fundamental principle that muscular loads, transmitted and modulated by the thorax, must match wing load requirements; what then is the thoracic modulation that ensures load-matching is satisfied? This identified thoracic modulation – and any associated elastic effects – would be a data-driven description of actual thoracic behaviour. This formulation is a true thoracic inverse problem. In simple terms, it attempts the identification of a black box modulator (the thorax) given both inputs (muscular loads and strains) and outputs (wing motion). In precise terms, neither muscular strain nor wing motion are true input or output parameters, as they are coupled by the muscular stretch-activation dynamics (Fig. 1C). This inverse problem corresponds more precisely to identifying properties of the thoracic wing/muscle feedback loop based on scattered observable properties within the loop. To the best of our knowledge, this approach has not previously been utilised for studying insect flight motors.

### 2.2. Sources of data

Attempting to solve the inverse problems formulated in §2.1 necessitates a degree of meta-analysis, as there is no single source or single methodology that provides all required data. Sources of data used in this study are tabulated in Table 1; and the associated datasets are illustrated in Fig. 2. These data sources may be specific to *D. Melanogaster*, or may involve generalisation from other species. They also vary in methodology: *in vivo* studies, *ex vivo* studies, and computational and experimental models are all utilised. Under a categorisation based on datasets (**1**) and (**2**) of §2.1, datasets utilised in this study are as follows.

(**1**) **Wing dataset**. Wingbeat kinematics for *D. melanogaster* are sourced from the published results of Beatus and Cohen [11] and Ben-Dov and Beatus [59], as well as additional in-house data (Maya *et al*. [63]). The synthetic kinematics used in the similitude model of Dickinson *et. al* [22] are also sourced for comparison. This kinematic dataset is shown in Fig. 2, scaled to a common peak-to-peak stroke amplitude (155°). Aerodynamic data for *D. melanogaster* is sourced from computational and experimental analyses showing appropriate Reynolds number (~100) and realistic wing planform. Two subcategories of aerodynamic data are distinguished. (**i**) (**i**) Data for synthetic wing kinematics, with zero elevation angle [22,26,27]. These results are characterised by presence of dual peaks, in both lift and drag, associated with wake capture and rotational circulation. (**ii**) Data for experimentally measured wingbeat kinematics, derived from motion tracking of *D. melanogaster* [17,23–25,28]. These do not show force peaks at stroke-reversal, but instead, asymmetric aerodynamic profiles with dominant lift and drag peaks in the middle of the upstroke. Existing studies have elucidated the effect of certain wingbeat kinematic parameters on the wingbeat aerodynamics – *e.g.*, the elevation angle [62,64,65], and stroke-incidence phase [22,26] – but as of yet there is no complete explanation for why the load waveforms generated by measured and synthetic kinematics differ so notably. All of these wing data sources are typically associated with slightly different wingbeat parameters (stroke amplitude, frequency, *etc.*). For consistency, we scale both kinematic and aerodynamic data to a unified set of *D. melanogaster* parameters, as given in Table 2. The details of this scaling process are given in the Supporting Information.

(**2**) **Muscular dataset.** Muscular load-strain profiles, for insect asynchronous flight muscles extracted from a range of species, are sourced from *ex vivo* studies [29–31]. The extent of muscular data applicable to the analysis of *D. Melanogaster* is limited: results from synchronous muscles [66,67] are not necessarily generalisable; and results specific to *D. Melanogaster* [30,31] show strain amplitudes significantly lower than in free flight [9]. The work of Josephson [29] is the basis for muscular analysis in this study: it provides empirical estimates of the load generated by an asynchronous flight muscle from a beetle, *Cotinus mutabilis*, under a range of large strain amplitudes. These results can be used to estimate muscular loads in *D. melanogaster* by appropriate scaling based on *D. melanogaster* muscle mass and cross-sectional area [30,31,47] and/or wing load and power requirements. *In vivo* muscular strain amplitudes and waveforms can be estimated based on high-speed video microscopy [9] and X-ray diffraction results [5].

**Table 1:** Data sources for thoracic inverse problem. Dataset parameters are: data sampling frequency $f_s$; Reynolds number $Re$; stroke angle peak amplitude $\hat{\phi}$; muscular strain peak amplitude $\hat{\varepsilon}$.

| Source | Ref. | | Conditions |
|---|---|---|---|
| **Wingbeat kinematic data** | Ref. | $f_s$ | Description |
| Maya *et al.* (2022) | [63] | 20 kHz | free flight |
| Ben-Dov and Beatus (2020) | [59] | 20 kHz | free flight |
| Beatus and Cohen (2015) | [11] | 8 kHz | free flight |
| Dickinson *et al.* (1999) | [22] | | synthetic model |
| **Aerodynamic data** | Ref. | $Re$ | $\hat{\phi}$ | Description |
| *Subcategory (i): synthetic kinematics* | | | | |
| Dickinson *et al.* (1999) | [22] | 136 | 80° | Experimental similitude |
| Sun and Tang (2002) | [27] | 136 | 72.5° | 3D CFD |
| Ramamurti and Sandberg (2002) | [26] | 136 | 80° | 3D CFD |
| *Subcategory (ii): biological kinematics* | | | | |
| Muijres *et al.* (2014) | [17] | not spec. | 67° | Experimental similitude |
| Meng *et al.* (2015) | [23] | 112-122 | 67-76.5° | 3D CFD (3x) |
| Meng *et al.* (2017) | [25] | 105 | 71° | 3D CFD |
| Shen *et al.* (2018) | [24] | 77-108 | 67.5-75° | 3D CFD (5x) |
| Yao and Yeo (2018) | [28] | 115 | 70° | 3D CFD |
| **Experimental muscular data** | Ref. | $\hat{\varepsilon}$ | Description |
| Josephson *et al.* (2000) | [29] | 1-4% | *Cotinus mutabilis*, basalar |
| Swank (2011) | [30] | 0.3% | *D. melanogaster*, DVM and DLM |
| Wang *et al.* (2011) | [31] | 0.3% | *D. melanogaster*, DVM and DLM |

**Table 2:** Unified parameter set for *D. melanogaster*

| Parameter | Source | Value | Scaling |
|---|---|---|---|
| *Stroke parameters:* | | | |
| air density ($\rho$) | [68] | 1.2 kg/m$^3$ | |
| wing mass ($m_w$) | [61] | 2.7 µg | |
| wingbeat frequency ($n$) | [69] | 218 Hz | |
| stroke peak amplitude ($\hat{\phi}$) | [11,50] | 77.5° | |
| *Single-wing parameters:* | | | |
| wing length ($R$) | [69] | 2.39 mm | |
| wing area ($S$) | [69] | 1.78 mm$^2$ | $0.376R^2$ |
| wing dim'less 2$^{nd}$ mom. of area ($\hat{I}$) | [69] | 0.35 | 0.35 |
| wing stroke mom. of inertia ($I_\rho$) | calc. | 5.40 µg mm$^2$ | $m_w \hat{I} R^2$ |
| wing aero. ref. pt. length ($r_a$) | [69] | 1.53 mm | $0.7R$ |

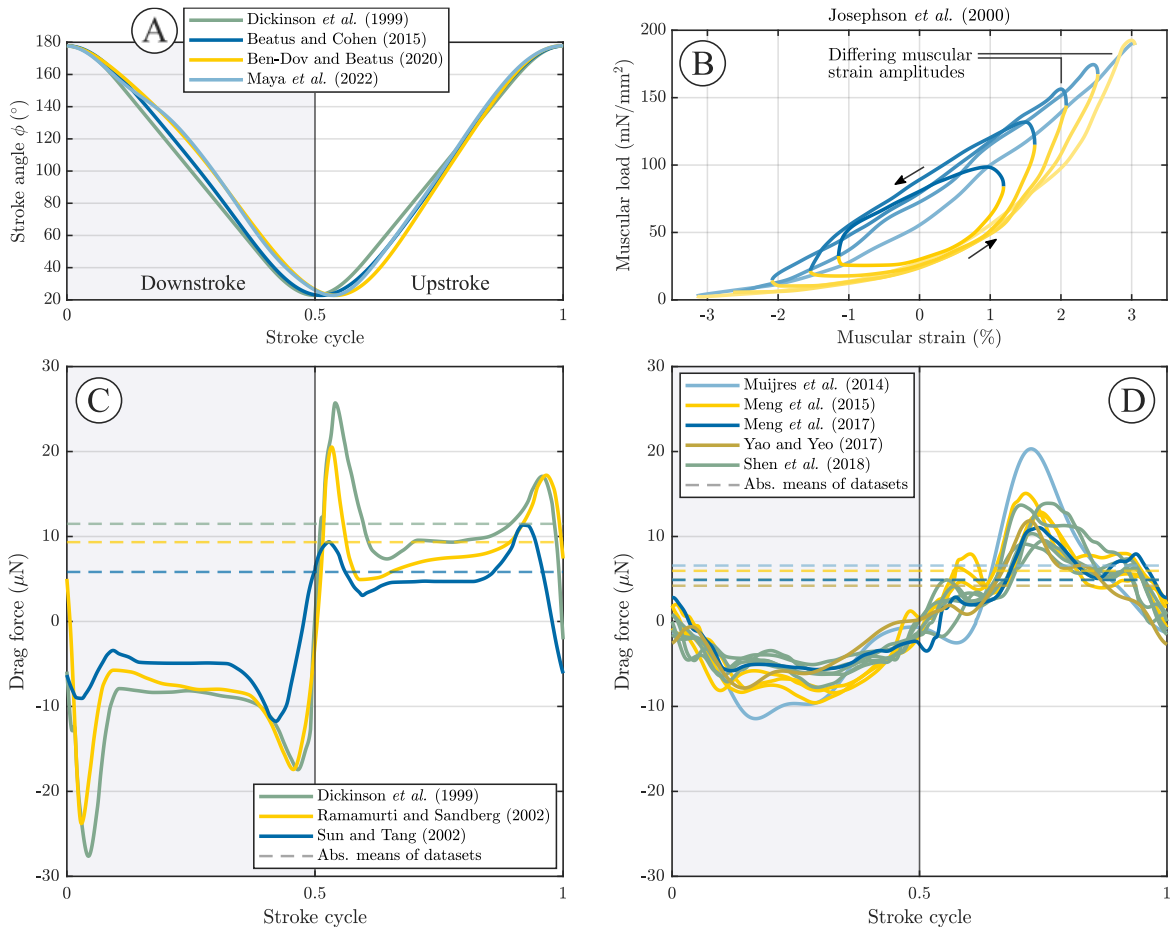

**Figure 2:** Key quantitative data sources for the thoracic inverse problem in *D. melanogaster*. (**A**) Wingbeat kinematics, as per [11,22,59,63], indicating the envelope (range) of wingbeat kinematics recorded. (**B**) Muscular load-strain data, as per [29]. (**C-D**) Wing drag force data, under classical synthetic (**C**) [22,26,27] and biological (**D**) [17,23–25,28] wingbeat kinematics.

**2.3. Dynamic and kinematic transmission**

§2.1 presented two inverse problems for identifying properties of the thoracic structure as part of the flight motor. The thoracic structure transmits and modulates the interaction between muscular load and strains, and wingbeat motion. This transmission takes on two coupled roles: (**i**) a kinematic role; referring to the transmission of displacements (strain, motion); and (**ii**) a dynamic role; referring to the transmission and modulation of load (stress, force, moment). In (**i**), the thorax connects the wing and flight muscles via some mechanical linkage, or kinematic chain [4,70]. This linkage allows muscular strain to generate wingbeat motion, and transmits muscular loads to the wing, via the mechanical advantage (or, disadvantage) afforded by the linkage. The presence of this kinematic linkage is a necessary property of the flight motor, and does not presuppose any elastic energy storage within the system. In (**ii**), the dynamic role, the thorax may do more than simply transmit load through a kinematic linkage. It may modulate these loads via elastic, inertial, or dissipative effects. In a description of the propagation of loads through the flight motor, these effects manifest as additional thoracic loads, dependent on muscular strains and wingbeat motion. To represent kinematic and dynamic transmission processes mathematically, we make two initial assumptions regarding the system behaviour, based on observational results.

Assumption (**1**). We take the wing stroke angle – as defined in a horizontal stroke plane – is taken to be the primary wing degree of freedom (DOF) for flight motor power consumption. Wing pitch kinematics may be attributed to passive elastic elements [11,61], and wing elevation kinematics are not thought to be energetically impactful [62]. It follows that the wing translational drag force – the force acting within the stroke plane – is the primary source of aerodynamic power consumption. If alongside this assumption, we take the tensile loads and strains on the flight muscles to be the primary DOF for muscular action – the accepted understanding of the operation of flight muscles [71–73] – then a single-DOF load transmission pathway can be isolated: from the stroke angle ($\phi$) to tensile muscular strains; $\varepsilon_i$, for dorsoventral ($i =$ DVM) and dorsolongitudinal ($i =$ DLM) muscles.

Assumption (**2**). We initially assume a one-to-one linear relation between muscular strains and wing stroke angle ($\varepsilon_i \propto \phi$). This is associated with a parallel-elastic actuation (PEA) model of the flight motor [74], in which thoracic elasticity acts primarily in parallel with the musculature, and thus, muscular strains and the wing stroke angle are perfectly in phase. We will examine the effect of certain nonlinearities in this relation later, but experimental evidence from both X-ray diffraction and vibrometer studies indicates that the relationship is largely linear and in phase [5,20]. If this is the case, then thoracic dynamic loads (elastic, inertial, dissipative) are solely responsible for differences in waveform between flight muscle load generation and wing load requirements.

Under these assumptions, the kinematic and dynamic effects of the thorax may be expressed as the following functional equations [75], in the time ($t$) domain:

*Dynamic transmission and modulation:*

$$\underbrace{M_{\text{iner}}(\phi(t)) + M_{\text{aero}}(\phi(t))}_{\substack{\text{wing inertial and aerodynamic} \\ \text{moments, dependent on wing} \\ \text{kinematics}}} = \underbrace{M_{\text{thorax}}(\phi(t))}_{\substack{\text{modulation of unknown} \\ \text{thoracic moments at} \\ \text{wing root}}} + \sum_{i \in \begin{bmatrix} \text{DVM} \\ \text{DLM} \end{bmatrix}} K_i \cdot \underbrace{F_i(\varepsilon_i(\phi(t)))}_{\substack{\text{muscular forces,} \\ \text{dependent on} \\ \text{muscular strain}}};$$

(1)

*Kinematic transmission:*

$$\underbrace{L_i}_{\substack{\text{muscle rest} \\ \text{length}}} \cdot \underbrace{\varepsilon_i(\phi(t))}_{\substack{\text{muscular} \\ \text{strains}}} = \underbrace{K_i}_{\substack{\text{stroke-strain} \\ \text{transmission} \\ \text{constants}}} \cdot ( \underbrace{\phi(t)}_{\substack{\text{wing stroke} \\ \text{angle}}} - \underbrace{\phi_{0,i}}_{\substack{\text{muscle resting} \\ \text{stroke angle}}} ).$$

Without loss of generality, the thoracic effective load $M_{\text{thorax}}$ is taken as a moment about the wing root. The identification of this thoracic load is the objective of inverse-problem solution. Eq. 1 is a set of functional equations: inertial, aerodynamic, and thoracic effective loads ($M_{\text{aero}}$, $M_{\text{iner}}$, $M_{\text{thorax}}$), as well as muscular loads ($F_i$) and strains ($\varepsilon_i$) are defined as functionals – functions of functions. These loads and strains are generated via complex, nested, dependencies on their dependent functions ($\varepsilon_i(\phi(t))$, $\phi(t)$): dependencies that may require information on the entire cumulative history of these dependent functions, and which encode a range of physical phenomena (aerodynamics, structural mechanics, *etc.*)

Simulating the flight motor under general conditions, via Eq. 1, requires the ability to compute all the functionals defined in Eq. 1 for arbitrary input. Crucially, however, this ability is not required for inverse-problem solution. Given data on input functions, *e.g.,* wingbeat kinematics, $\phi(t)$, the only information that is required is the functional output at these prescribed input functions, *e.g.*, aerodynamic loads for $\phi(t)$. This leads to a simple relation, without functional dependency. To define this relation, however, several further processes need to be defined.

## 2.4. Aerodynamic and inertial loads

Wing aerodynamic and kinematic data must be translated into aerodynamic and inertial moments about the wing root ($M_{\text{aero}}$, $M_{\text{iner}}$). We begin with single-wing drag forces, $F_D(t)$, and stroke angle kinematics, $\phi(t)$, for our model *D. melanogaster*, as per the data scaling process set out in the Supporting Information. The single-wing aerodynamic drive moment is computed as:

$$M_{\text{aero}}(t) = r_a F_D(t), \qquad (2)$$

where $r_a$ is the location of the wing aerodynamic centre along the span, which we estimate as $r_a = 0.7R$ for wing spanwise length $R$ [26,69] (Table 2). Note that recent studies indicate that the location of the wing aerodynamic centre may vary over the wingbeat cycle [76,77], but without detailed data, an assumption of fixed location is retained. The single-wing inertial drive moment is computed as:

$$M_{\text{iner}}(t) = m_{\text{w}} \hat{I} R^2 \ddot{\phi}(t). \qquad (3)$$

where $m_{\text{w}}$ is the wing mass and $\hat{I}$ the wing second moment of area. This computation assumes that the stroke angle is the only wing DOF directly actuated (§2.3); and that the wing shows approximately uniform mass density. In addition the inertia associated with the stroke angle ($\phi$) is assumed to be independent of wing flexion, wing pitch rotation, and wing elevation rotation.

## 2.5. Muscular loads

To solve the thoracic inverse problem for consistent thoracic modulation (Fig. 1), muscular loads must be integrated into the transmission functionals (Eq. 2). To construct an appropriate muscular functional output given data in the literature, the following process is carried out.

(**i**) **Scaling.** Muscular loads are taken to scale linearly with muscular cross-sectional area, with load per unit cross-sectional area given in existing studies of muscular behaviour (Table 1, Fig. 2):

$$F_i(\varepsilon_i) = A_i u_i(\varepsilon_i), \qquad i \in [\text{DVM}, \text{DLM}] \qquad (4)$$

for muscular cross-sectional area $A_i$ and load-per-area as a function of strain $u_i(\varepsilon_i)$. We assume that load-per-mass data from the asynchronous muscles of other species are the same for the flight muscles of *D. melanogaster*; and that effects of variation in temperature are negligible [29,71,78].

(**ii**) **Duplication.** In *D. melanogaster*, two flight muscles act in synchrony: the dorsoventral (DVM) and dorsolongitudinal (DLM) muscles. We neglect the effect of the flight motor steering muscles, which are known to represent <3% of the total flight muscle mass in other Dipterans [3]. The operation of this dual-muscle system must be reconstructed from single-muscle load-strain data [29–31], and the results of *in vivo* observation of dual-muscle operation [5,9]. Based on observational data, we assume the DVM and DLM show approximately equivalent action: equivalent length and total cross-sectional area [3,9,33]; equivalent strain amplitude at 180° phase offset [5,9]; and load-per-area profiles $u_i(\varepsilon_i)$ that are equivalent but act at opposite strain proportionality. The latter is based on the principle that drag and inertial loads should be approximately equal during upstroke and downstroke (Fig. 2). The resulting dual-muscle model (the summation over $i$ in Eq. 1) is symmetric about the midstroke point, as shown in Fig. 3A. Formally:

$$\begin{aligned} A_{\text{DLM}} &= A_{\text{DVM}}, \qquad K_{\text{DLM}} = -K_{\text{DVM}}, \\ u_{\text{DLM}}(\varepsilon_{\text{DLM}}) &= u_{\text{DVM}}(-\varepsilon_{\text{DVM}}). \end{aligned} \qquad (5)$$

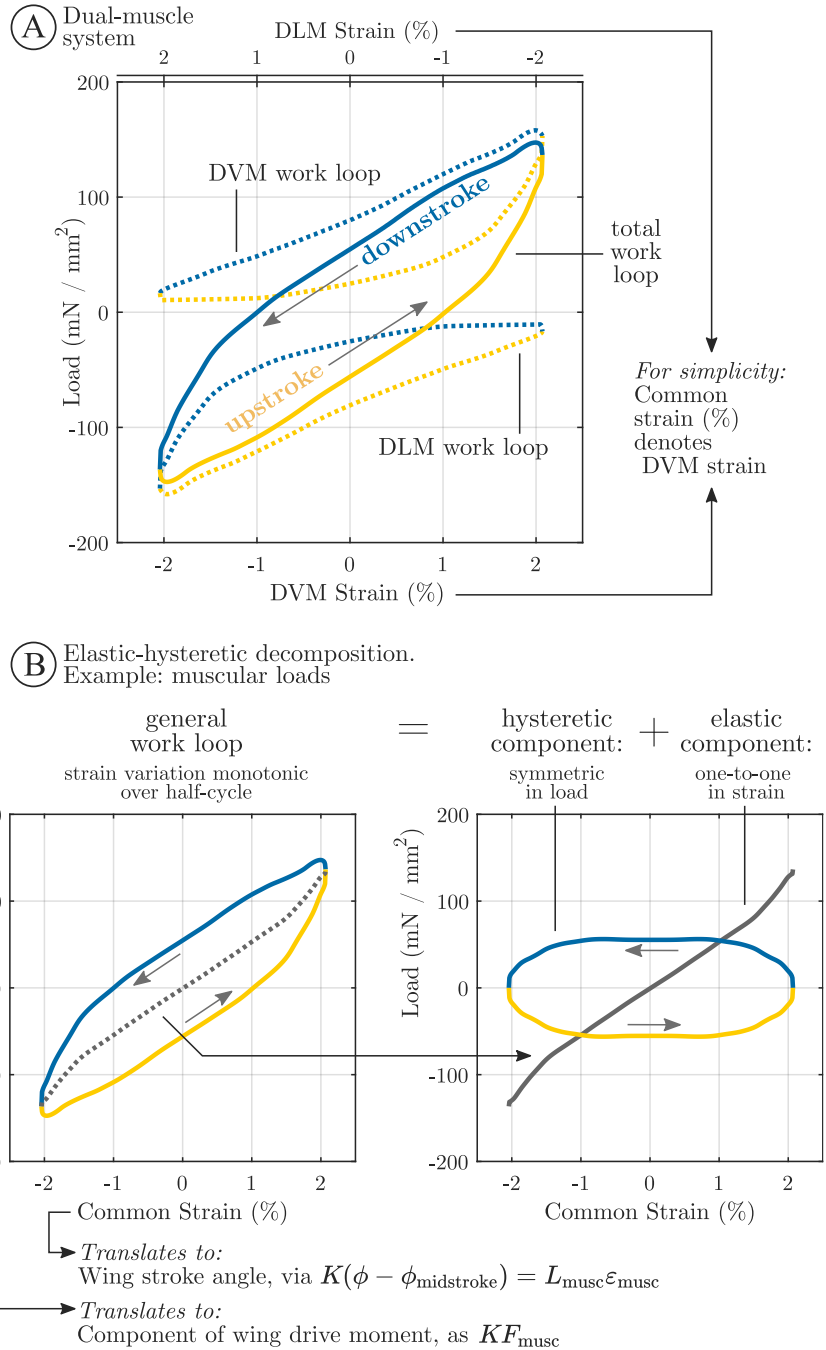

**Figure 3:** Work-loop properties of the flight motor. (**A**) Dual muscle configuration in the flight motor, as modelled by single-muscle data. The resulting dual-muscle load represents the summation over $i$ in Eq. 1. (**B**) The operation of elastic-hysteretic decomposition, performed on dual muscle loads as an example. Note that work-loop principles apply both to visualisations in strain and load; and those in stroke angle and moment. The relationship between these pairs is given by Eq. 6.

### 2.6. Transmission solution

With loads defined as per §2.4-2.5, the thoracic inverse problem may be defined on the flight motor system expressed in terms of calculable quantities:

*Thoracic dynamic transmission and modulation:*
$$M_{\text{iner}}(\phi(t)) + M_{\text{aero}}(\phi(t)) = M_{\text{thorax}}(\phi(t)) + KF_{\text{musc}}\bigl(\varepsilon_{\text{musc}}(\phi(t))\bigr). \tag{6}$$

*Thoracic kinematic transmission:*

$$L_{\text{musc}} \cdot \varepsilon_{\text{musc}}(\phi(t)) = K \cdot (\phi(t) - \phi_{\text{midstroke}}).$$

*Dual-muscle system:*

$$F_{\text{musc}} = K m_{\text{musc}} u_{\text{musc}},$$
$$u_{\text{musc}} = u_{\text{single musc}}(\varepsilon_{\text{musc}}) - u_{\text{single musc}}(-\varepsilon_{\text{musc}}),$$
$$K_{\text{musc}} = K_{\text{DLM}} = -K_{\text{DVM}}, \qquad L_{\text{musc}} = L_{\text{DLM}} = L_{\text{DVM}},$$
$$A_{\text{musc}} = A_{\text{DLM}} + A_{\text{DVM}}.$$

Transmission functionals are defined as per Eq. 1. For periodic wingbeat motion, we utilise a stroke cycle parameter $x \in [0,1]$, representing location in the stroke cycle. Eq. 6 thus yields the simple inverse-problem solution for $M_{\text{thorax}}$:

$$M_{\text{thorax}}(x) = M_{\text{iner}}(x) + M_{\text{aero}}(x) - N u_{\text{musc}}(x) \tag{7}$$

for some muscular scale factor $N$. We will estimate $N$ during the analysis process by matching the net work production of the muscles to the net work requirement associated with wingbeat motion.

## 2.7. Work loop analysis and the elastic-hysteretic decomposition

An information-rich aspect of the loads, $M(x)$, in Eq. 7 is their associated work loops: the loop traced out by a visualisation of load, $M$, against stroke angle $\phi$ (*cf.* Fig. 3B) [29,79]. The transmission relation, Eq. 7, can be represented as a work loop equation of motion [74,80]. For load $M(x)$, splitting the work loop in $M$ against $\phi$ into upper and lower boundaries – $M^+(\phi)$, associated with the downstroke, and $M^-(\phi)$, associated with the upstroke – generates the work-loop equations:

$$M^{\pm}_{\text{thorax}}(\phi) = M^{\pm}_{\text{iner}}(\phi) + M^{\pm}_{\text{aero}}(\phi) - N u^{\pm}_{\text{musc}}(\phi). \tag{8}$$

Going further, these work loops, denoted generally, $M^{\pm}(\phi)$, can be decomposed into components which are purely elastic (conservative) and purely hysteretic (dissipative), as per Fig. 3B. Formally:

$$M^{\pm}(\phi) = M_{\text{elastic}}(\phi) \pm M_{\text{hysteretic}}(\phi) \tag{9}$$

where $M_{\text{elastic}}(\phi)$ purely elastic, defining the mean value of a work loop as a function of $\phi$, and $M_{\text{hysteretic}}(\phi)$ is purely hysteretic, defining the half-width (in $M$) of the loop. These two components broadly correspond to structural elastic and structural dissipative effects, respectively. Structural elasticity is necessarily a conservative force, $M = k(\phi)\phi$, and contributes entirely to $M_{\text{elastic}}(\phi)$. Damping, if described as $M = d(\dot{\phi})\dot{\phi}$, contributes entirely to $M_{\text{hysteretic}}(x)$, if the upstroke and downstroke kinematics are symmetric. If not, or if there is positional dependency $M = d(\phi, \dot{\phi})\dot{\phi}$, then some aliasing will be observed in $M_{\text{elastic}}(\phi)$. Inertial effects, including muscular inertia for the muscular work loop, wing inertia for the wing loop, *etc.*, are necessarily defined as $M = m(\phi)\ddot{\phi}$, and thus may alias onto both profiles. This decomposition thus allows a basic identification of structural and dissipative mechanisms active within the thorax – though inertial aliasing effects must be kept in mind. Notably, the decomposition (Eq. 9) applied to the transmission (Eq. 8) also allows independent transmission relationships in elastic and hysteretic components to be formulated, that is:

$$M_{\text{thorax,elastic}}(\phi) = M_{\text{iner,elastic}}(\phi) + M_{\text{aero,elastic}}(\phi) - N u_{\text{musc,elastic}}(\phi),$$
$$M_{\text{thorax,hysteretic}}(\phi) = M_{\text{iner,hysteretic}}(\phi) + M_{\text{aero,hysteretic}}(\phi) - N u_{\text{musc,hysteretic}}(\phi), \tag{10}$$

meaning, *e.g.*, that elastic properties do not need to be precisely known in order to study the transmission of dissipative loads. We will return to this particular property in §4.

## 2.8. Metrics of power consumption

Finally, as quantitative measures of flight motor power consumption, several mean mechanical power consumption metrics are available. The choice is associated with differing treatments of flight motor negative work – work done by the wing on the flight motor [74,80,81]. Two key metrics may be defined: (**a**) the absolute power, $\overline{P}_{abs}$, with components of negative work requiring equivalent resistance power; and (**b**) positive-only power, $\overline{P}_{pos}$, with components of negative work dissipated perfectly. In our flight motor context, these metrics may be computed as:

$$\overline{P}_{abs} = \frac{1}{T}\int_0^T \left|\dot\phi K F_{musc}(t)\right| dt.$$
$$\overline{P}_{pos} = \frac{1}{T}\int_0^T \dot\phi K F_{musc}(t)\left[\dot\phi K F_{musc}(t) > 0\right] dt, \quad (11)$$

Clearly, $\overline{P}_{abs} \geq \overline{P}_{pos}$. The positive-only power has been used in previous waveform-based energetic analyses of flight motor behaviour [48,50,82]; but does represent an optimistic assessment of the role of dissipation within the flight motor. The absolute power has been used in other biomechanical contexts, such as bipedal walking and limb movement [81,83,84]. With either of these metrics, we can quantify the effect of thoracic elasticity on flight motor power consumption. Computing a hypothetical power requirement without elasticity ($\overline{P}_{[\ldots],\,w/o\,elast.}$), and one with elasticity ($\overline{P}_{[\ldots],\,with\,elast.}$), we can estimate the mechanical power *savings* associated with flight motor elasticity as:

$$\eta_{abs} = \frac{\overline{P}_{abs,\,w/o\,elast.} - \overline{P}_{abs,\,with\,elast.}}{\overline{P}_{abs,\,w/o\,elast.}},$$
$$\eta_{pos} = \frac{\overline{P}_{pos,\,w/o\,elast.} - \overline{P}_{pos,\,with\,elast.}}{\overline{P}_{pos,\,w/o\,elast.}}. \quad (12)$$

The role of elasticity in generating these energy savings involves storing and releasing negative work [46,48,74]. This can be seen in terms of the work loop formulation of §2.4. If the thoracic modulation is purely elastic, $M^{\pm}_{thorax}(\phi) = M_{elastic}(\phi)$, then the elastic modulation(s) which minimise $\overline{P}_{abs}$ and $\overline{P}_{pos}$ simultaneously are given by the elastic-bound conditions [74]:

$$M^{-}_{iner}(\phi) + M^{-}_{aero}(\phi) \leq -M_{elastic}(\phi) \leq M^{+}_{iner}(\phi) + M^{+}_{aero}(\phi). \quad (13)$$

Note that, in general, this defines not just one elastic profile, but a continuous range: any sign-flipped elastic profile lying within the bounds of the inelastic work loop ($M^{\pm}_{iner}(\phi) + M^{\pm}_{aero}(\phi)$) is optimal in $\overline{P}_{abs}$ and $\overline{P}_{pos}$. We note, for formal validity, that the work loops in our dataset satisfy the validity conditions associated with the elastic-bound conditions (closed loops no more than bivalued at any $\phi$, loops purely dissipative) [74]. We study the implications of these conditions in §3.2.

## 3. Identification by optimality
### 3.1. Inertial load waveforms imply strain-hardening nonlinearities

The optimality inverse problem described in §2.1 involves the identification of the thoracic modulation that optimises some metric of the muscular load required to generate the observed wingbeat motion, assuming the muscles can provide load at *any* waveform. Two selectable aspects of this optimisation process are (**i**) the allowable forms of modulation (elastic, dissipative, inertial; linear, nonlinear; *etc.*), and (**ii**) the selected optimality metric (power, load; mean, peak; *etc.*). Evidence for thoracic resonance (§1) would imply that it is specifically thoracic elasticity which minimises muscular power requirements via absorption of wing inertial loads.

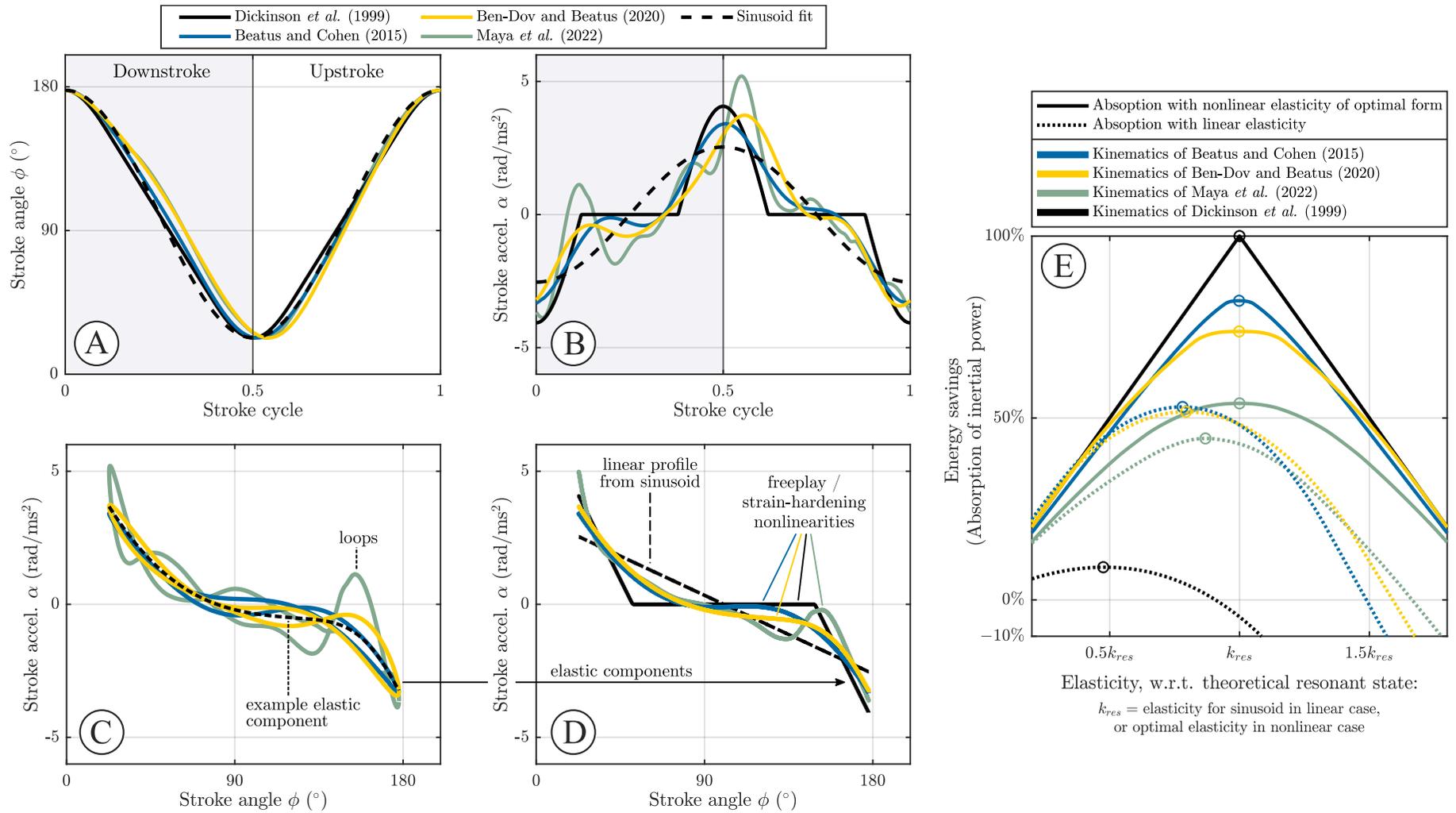

**Figure 4:** Effect of stroke waveform on wing inertial loads and resonant effectiveness in *D. melanogaster*. (**A**) Wing stroke angle kinematics: synthetic, biological, and sinusoidal fit. (**B**) Stroke angular acceleration ($\propto$ inertial load) profiles. (**C-D**) State-space view of stroke angular acceleration profiles, with elastic-hysteretic decomposition generating the required elastic profile for optimal absorption of inertial loads. (**E**) Energetic effect of elasticity: actuator energy consumption, using metric $P_{(b)}$ (§2.8) for this undamped system, that can be achieved with linear and optimal nonlinear elastic profiles.

Page 13 of 34

Consider then a simple initial analysis of thoracic resonant modulation, in which wing inertial loads are matched with overall flight motor elasticity, independent of aerodynamic effects (Fig. 4). Note that this overall flight motor elasticity includes both the elasticity of the thoracic structure, and implicitly, the intrinsic elasticity of the primary flight muscles. If the wing stroke kinematics of *D. melanogaster* were purely sinusoidal, then inertial loads ($M_{\text{iner}} \propto \ddot{\phi}$) and elastic loads ($M_{\text{thorax}} \propto \phi$) would be directly proportional: $\ddot{\phi} \propto \phi$. Both loads would appear as straight lines in the state-space $M$-$\phi$, and it would follow that perfect absorption of inertial loads could be achieved via a linearly elastic thorax ($M_{\text{thorax}} = k_{\text{thorax}}\phi$). However, the wing stroke kinematics of *D. melanogaster*, and certain other insects, are not purely sinusoidal. Instead, they are commonly thought to resemble smoothed sawtooth or triangle waveforms [22,47,62,64,85]. This difference is small when observed as $\phi(t)$ (Fig. 4A) but has significant implications for thoracic resonance. When observed as the derivative $\ddot{\phi}(t)$, triangle-type waveforms show a pulsed profile, with periods of near-zero acceleration associated with the area of constant-velocity motion about the midstroke (Fig. 4B). When observed as a loop in the space $\ddot{\phi}$-$\phi$ (or, equivalently, as the work loop $M_{\text{iner}}$-$\phi$), the optimal profile for the thoracic elasticity, $M_{\text{thorax}}(\phi)$, is nonlinear (Fig. 4C). If the $M_{\text{iner}}$-$\phi$ profile is a single line (no dissipation) then this profile is exactly and uniquely the optimal elastic function $M_{\text{thorax}}(\phi)$, and this elasticity completely absorbs all inertial load and power requirements. If the $M_{\text{iner}}$-$\phi$ profile has some hysteretic component (§2.6), then an optimal elastic function $M_{\text{thorax}}(\phi)$ can be isolated as the elastic component of the $M_{\text{iner}}$-$\phi$ profile. The remaining hysteretic component is a muscular load requirement which cannot be absorbed by thoracic elasticity. For the synthetic kinematics of [22], the $M_{\text{iner}}$-$\phi$ profile, and thus the optimal $M_{\text{thorax}}(\phi)$, is exactly a freeplay nonlinearity (Fig. 4D). For measured biological kinematics, a cubic-like nonlinearity is observed (Fig. 4D). In both cases, the elasticity may be characterised as strain-hardening.

The key implication of this result is that the thoracic elasticity that optimally absorbs inertial loads (independent of aerodynamic loads) is strain-hardening, with a strong resemblance to freeplay nonlinearity. This contrasts with current analyses-based studies of the energetics of thoracic elasticity, which typically assume linear thoracic elasticity [41–43,86]. This assumption may not be safe: Fig. 4E shows the result using linear elasticity to absorb the inertial loads associated with biological stroke kinematics: resonant efficiencies are markedly reduced. It follows that biomimetic MAVs attempting to replicate biological stroke kinematics and biological thoracic resonance may be well-suited to similar strain-hardening nonlinearities. Cubic-type nonlinear elasticities have already been considered for use in FW-MAVs [87], and this result provides a biomimetic basis for such nonlinearities.

### 3.2. The landscape of thoracic elastic optimality

§3.1 considered an initial optimisation of thoracic elasticity based only on the absorption of inertial loads by elastic loads, independent of any aerodynamic loading. The principle of inertial load absorption still holds in an aerodynamically-damped environment; and the associated optimal elasticity is still relevant; but the situation is considerably more complex. Two effects significantly alter the context for elasticity optimisation. The first effect, is that aerodynamic load waveforms may interact with inertial load waveforms in non-trivial ways: aerodynamic loads may contain components, arising, *e.g.*, from aerodynamic added mass and vortex capture [22,88], and these components may alter the effective inertial load that is able to be absorbed by elasticity. The second effect is that states of resonant optimality in the flight motor system may become distinct, and mutually exclusive [43]. In the analysis of §3.1, there were single optimal states: elasticities which near-perfectly absorbed all inertial effects, and thereby led to zero load and power requirements associated with structural inertia. However, when aerodynamic damping (or, any other form of damping) is present in the system, the elasticities which optimise metrics of load and power consumption in the complete system may differ from each other

[43]. Even further, these optimal elasticities may be non-unique – that is, there may be a set of elasticities which all equally minimise, *e.g.*, the mean power consumption, $\overline{P}_{abs}$ or $\overline{P}_{pos}$. Indeed, theoretical analyses has shown [74,80] that minimising $\overline{P}_{abs}$ or $\overline{P}_{pos}$ necessarily leads to a non-unique optimum in a wide class of systems – with a few conditions, any nonlinear elasticity lying within bounds defined by the system work loop is optimal in $\overline{P}_{abs}$ or $\overline{P}_{pos}$ [74].

These two effects make the landscape of optimal thoracic elasticity complex. However, several salient features can be identified. Considering the minimisation of mean mechanical power consumption, $\overline{P}_{abs}$ or $\overline{P}_{pos}$, we analyse a flight motor dataset involving the experimentally-observed kinematics of Beatus and Cohen [11], Ben-Dov and Beatus [59], and Maya *et al.* [63] for inertial load computation, and the full aerodynamic dataset given in Fig. 2 (14 drag force profiles) for aerodynamic load computation. This process involves mixing kinematic data from one source with aerodynamic data from a different source in a meta-analysis process (§2.8). In doing so, we pay a price in inconsistency (*e.g.*, between inertial and aerodynamic models) in exchange for increased accuracy in individual model components (*e.g.*, the ability to utilise the most recent kinematic estimates) and the ability to quantify uncertainty (*e.g.*, via the envelopes defined by the full dataset). From the dataset of 33 combinations of kinematic and aerodynamic models, we calculate load requirement estimates, and identify:

(**i**) **Energy consumption**: $\overline{P}_{abs}$ and $\overline{P}_{pos}$ in the absence of thoracic elastic modulation (Fig. 5A). Values are in the range 9 μW to 21 μW for a single wing; corresponding to a full-insect mass-specific power of 18 W/kg to 42 W/kg (w.r.t. full body mass, taken as 1 mg); consistent with existing mass-specific power estimates [46]. Power requirements for aerodynamics based on biological kinematics are, on average, 25% smaller than those based on synthetic kinematics.

(**ii**) **Resonant energy savings**: the maximum reduction in $\overline{P}_{abs}$ and $\overline{P}_{pos}$ possible via an optimal thoracic elastic modulation (Fig. 5C). Our estimates are in the range of 0-19% ($\overline{P}_{abs}$) and 0-10% ($\overline{P}_{pos}$) for aerodynamics based on synthetic wingbeat kinematics; and 8-31% ($\overline{P}_{abs}$) and 4-18% ($\overline{P}_{pos}$) for aerodynamics based on biological wingbeat kinematics. These estimates are given by the elastic-bound conditions (§2.8) [74], and are related to the fraction of negative work in the wingbeat drive load requirement ($M_{iner} + M_{aero}$). The difference in maximum power reduction between synthetic and biological wingbeat kinematics is notable. The larger reductions available to biological kinematics does not in itself indicate that these kinematics are more efficient or effective overall (lift generation is not considered). It indicates rather that, for biological kinematics, the energetic effects of elasticity may be particularly important. This effect is due not only to the smaller mean aerodynamic power requirement associated with biological kinematics (Fig. 5A), but also, the fact that biological aerodynamic loads appear much more like classical damping loads (elliptical work loops) and thus interfere less with the absorption of inertial negative work (Figs. 2, 5B).

(**iii**) **Resonant elasticities**: the set of elasticities that ensure this maximum reduction in $\overline{P}_{abs}$ and $\overline{P}_{pos}$, which are also given by the elastic-bound conditions [74]: any sign-flipped elasticity, $-M_{thorax}$ bounded by the inelastic work loop (the loop of $M_{iner} + M_{aero}$) minimises both $\overline{P}_{abs}$ and $\overline{P}_{pos}$. Within this set of elasticities, the midline elasticity (the elastic component of the loop $M_{iner} + M_{aero}$) then ensures that peak load and power are also minimised [74]. This midline elasticity is thus the 'most optimal' in a certain sense – midline elasticities for our data-driven model are illustrated in Fig. 5. We observe again that these elasticities are strain-hardening: reinforcing the conclusions of our simplified analysis that strain-hardening elasticities are energetically optimal flight motor elasticities in *D. melanogaster*.

Returning to an overall view of the role of flight motor elasticity in *D. Melanogaster*, one further point may be noted. We may observe that the work loop midline elasticity is optimal in peak load and power, *cf.* (**iii**) above, However, if peak load and power are not limiting factors, then, by selecting a different elasticity from the non-unique optimum in mean power consumption, the load and power waveforms that the flight motor must provide (*i.e.*, the required $u_{\text{musc}}$) can be altered significantly. That is, flight motor elasticity allows the wingbeat load and power requirement waveforms to be tuned, or synchronised, to the optimal action of the flight musculature – while remaining optimal in overall power consumption. Figure 6 illustrates a simple example of this effect.

Starting with the midline elastic profile for the biological wingbeat kinematics (Fig. 5C), we study two different parameterised alterations of this profile (Fig. 6). The first alteration involves shifting the profile uniformly towards either the upper or lower boundaries of the inelastic work loop (the boundaries of the optimal zone, §2.8). The second alteration involves shifting the profile simultaneously to both boundaries, in the manner of a bistable elasticity. Formally, our alterations from midpoint profile, $M_{\text{elast,mid}}(\phi)$, to altered profile, $M_{\text{elast,alter}}(\phi, \lambda)$, as a function of alteration magnitude $\lambda$, are:

$$M_{\text{elast,alter}}(\phi, \lambda) = (1 + \lambda)W_i(\phi)\big(M_{\text{elast,mid}}(\phi) + M_{\text{ref}}\big) - M_{\text{ref}}, \qquad i \in [1,2]. \tag{14}$$

where $W_1(\phi)$ and $W_2(\phi)$ are the first and second alteration normal forms, $M_{\text{ref}} = 50$ nNm is a reference moment (the baseline for alteration), and $\hat{\phi}(\phi)$ is the normalised stroke angle:

$$\begin{aligned}
W_1(\phi) &= \sin^{\frac{3}{2}}\left(\pi\hat{\phi}(\phi)\right), \\
W_2(\phi) &= \sin\left(2\pi\hat{\phi}(\phi)\right)\sin^3\left(\pi\hat{\phi}(\phi)\right), \\
\hat{\phi}(\phi) &= \frac{\phi - \min(\phi)}{\max(\phi) - \min(\phi)},
\end{aligned} \tag{15}$$

Application of these alterations, within the boundaries of the optimal zone, leads to shifts in the computed distribution of flight motor load/power requirement over the stroke cycle. The first alteration (Fig. 6A) shifts load/power between upstroke and downstroke: in the limit case, leading to the one-way drive system of [74], in which the flight motor actuation is near-unidirectional. This result indicates that asymmetry in the DVM/DLM load waveforms can be consistent with energetic optimality in the flight motor: even if the wingbeat kinematics are symmetric, this does not necessitate symmetric muscular loading. The second alteration (Fig. 6B) shifts load/power between each half of the upstroke and downstroke: from the first and third quarter-stroke to the second and last, and vice versa. In *D. Melanogaster*, this power transfer would represent a preference for muscular loading immediately after high muscular extension (1st/3rd quarter stroke) vs. after low muscle extension (2nd/4th quarter stroke): a feature potentially related to muscular stretch activation [31].

*Overleaf:* **Figure 5:** Energetic requirements of the *D. melanogaster* wing under both inertial and aerodynamic loads. (**A**) Drive moments and power requirements in the time domain, indicating the contribution of inertial and aerodynamic loads, and the envelope of the data sources considered in this work. Note the regions of negative power, corresponding to energy that could be absorbed by an elastic element. (**B**) Drive moments in the displacement ($\phi$) domain: system work loops. (**C**) Energetic interpretation of the system work loops. The midpoint moment, *i.e.*, the moving average of the work loop over $\phi$, is a metric of the effective inertial load of the system, and thus, with a sign flip, the energetically-optimal resonant elasticity. However, this elasticity is not unique: any elasticity bounded by the work loop is also energetically optimal [74]. This permits a wide range of elasticities, but not, in this system, a linear elasticity, which lies outside the work loop.

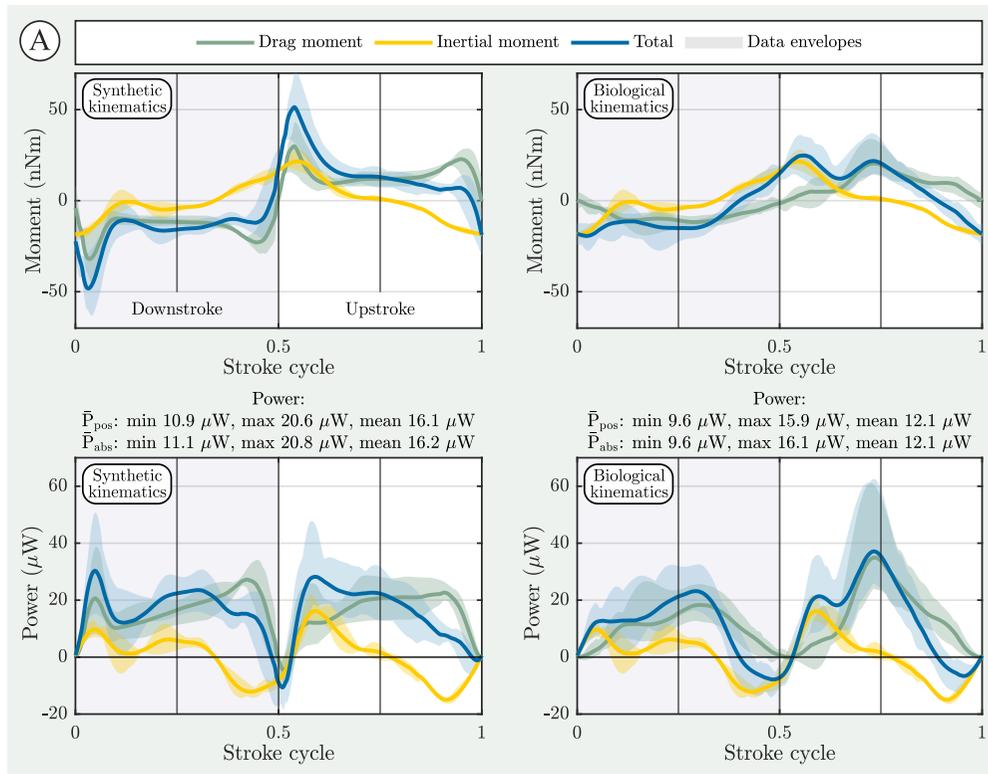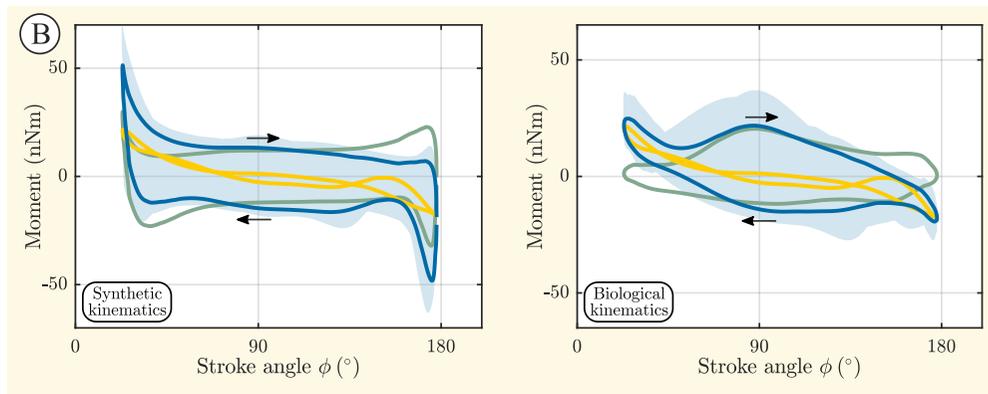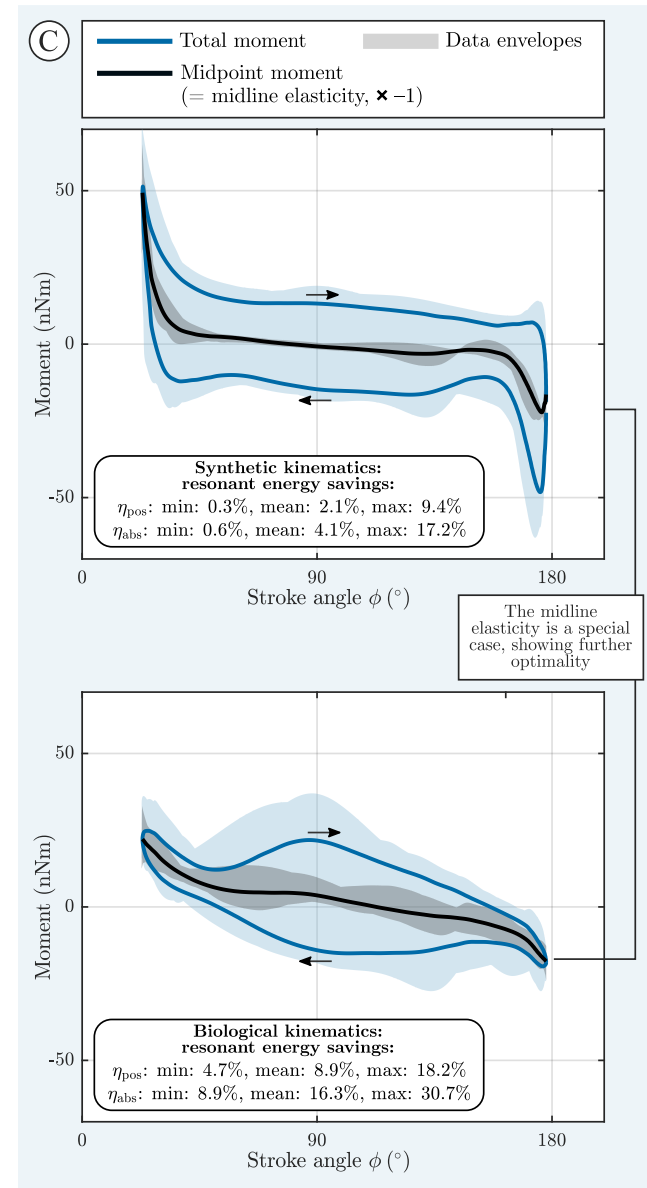

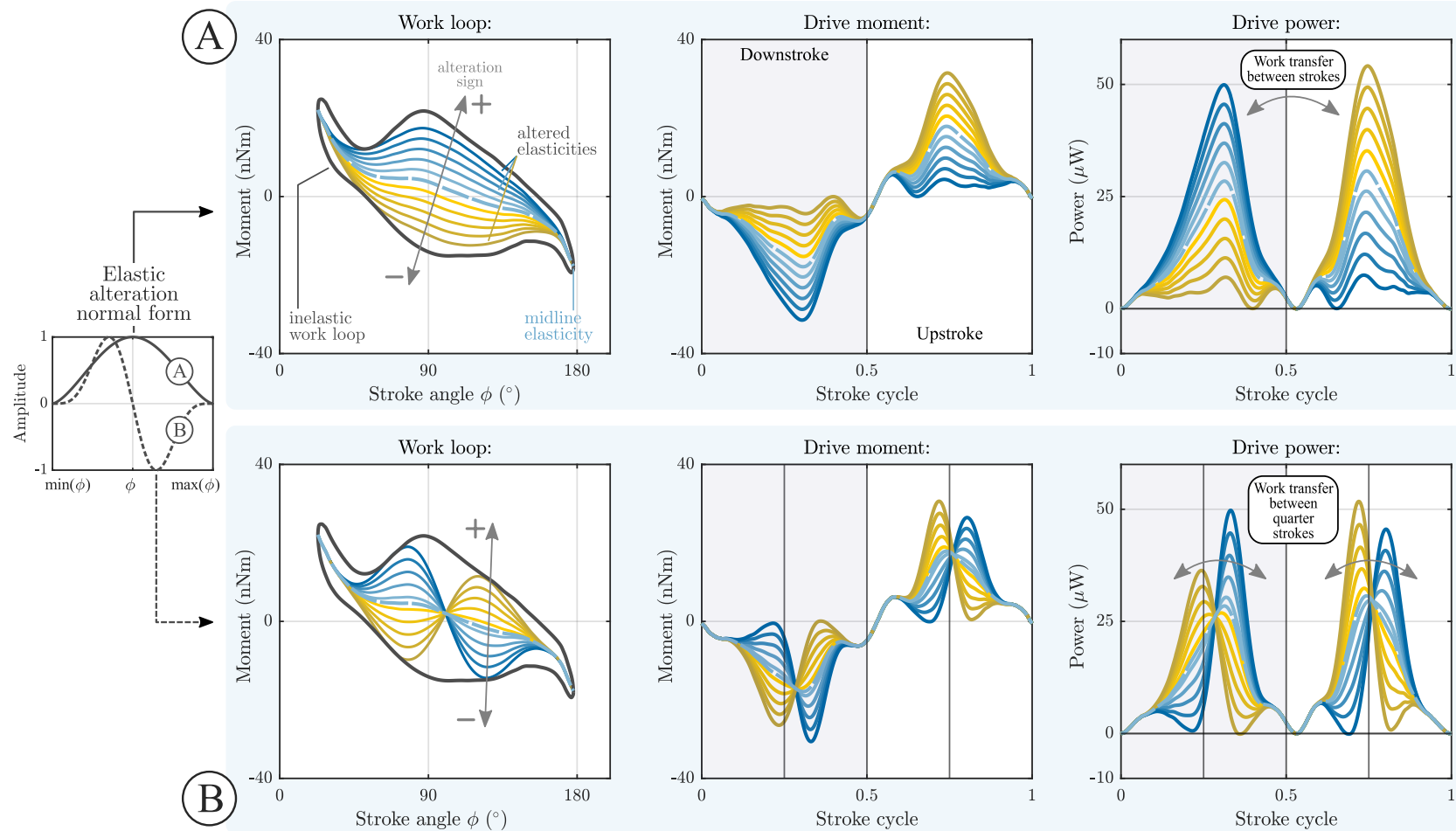

**Figure 6:** Example load and power waveform alteration achievable, at an energetically optimal state, by alteration of flight motor elasticity within the bounds of the optimal zone (the inelastic work loop). (**A**) First form of alteration: elasticity profiles and resulting drive load and power requirements. This alteration leads to work transfer between upstroke and downstroke, and leads in the limit case to the one-way drive system of [80]. (**B**) Second form of alteration: elasticity profiles and resulting drive load and power requirements. This alteration leads to work transfer between quarter-strokes.

# 4. Identification by consistency
## 4.1. Hysteretic musculature action follows an elliptical profile

The analysis of §3 followed the inverse problem for motor optimality (§2.1). By making assumptions about the energetic properties that the *D. melanogaster* flight motor optimises (*e.g.*, power consumption), we identified sets of appropriate thoracic elasticities that ensure optimality in these properties. Turning to the inverse problem for consistent transmission (§2.1), we can eliminate the need to make assumptions about the energetic priorities of the *D. melanogaster* flight motor. With data-driven models of both the operation of the main flight muscles (the DVM and DLM), and the wingbeat dynamics and aerodynamics, we can identify properties of the thorax that ensure physical consistency between flight muscle operation and wingbeat motion (*cf.* §2.6-§2.7).

This identification no longer rests on assumptions about optimality; but only on the accuracy of our source data and analysis framework. It is necessary to make approximations in order to construct an adequate muscular model for this process (§2.5). Note in particular that existing analyses of DVM/DLM muscular strain profiles in *D. virilis* indicate peak-to-peak strain amplitudes of between 2% and 5%, with a mean of 3.3% for the DVM and 3.5% for the DLM [9]. These strain amplitudes, which we take as representative of *D. melanogaster*, are significantly larger than those used in existing experiments into *D. melanogaster* flight muscle behaviour [30,31], and so we are forced to rely on data for the asynchronous flight muscle of a non-Dipteran species: a beetle, *C. mutabilis* [29], as per §2.2. The degree to which this data is generalisable to *D. melanogaster* is a source of uncertainty.

Given this uncertainty, we observe that there is a key feature of muscular load-strain data which appears highly generalisable: the hysteretic component of the muscular work loop (determining the net work generation of the muscle) is well approximated by an elliptical profile (Fig. 7A). When the hysteretic component of the loop is normalised over strain and load, this profile is well-approximated by a circle (Figs. 7B,C). This circular profile is present irrespective of the strain amplitude, and, indeed, of the muscular dataset – it may be observed in dual-muscle models constructed from other literature muscular data [30,31]. This point of commonality is important: in §4.2 we will utilise it to shed light on the behaviour of the *D. melanogaster* flight motor vis-à-vis its wingbeat kinematics.

## 4.2. Biological wingbeat kinematics are consistent with muscular load profiles; synthetic kinematics are not.

As per our transmission model in §2.3-2.7, in the flight motor of *D. melanogaster*, the loads generated by the musculature must match the loads required to generate the known wingbeat motion. Each of these loads may be defined in the form of work loops, which may be decomposed into elastic (midline) and hysteretic (width) components (§2.7). In our PEA model (§2.3), we may analyse the elastic and hysteretic components of these loops separately (Eq. 10). Clearly, an elastic component cannot influence a hysteretic component, and vice-versa: *e.g.*, altering the midline of the work loop does not alter its width. Here, load matching implies matching each of these two work loop components separately (Eq. 10). Any difference between muscular elasticity, and the elastic component of the wingbeat load requirement must be the result of thoracic elastic and/or inertial effects. Any difference between the muscular hysteretic loop (net work output) and the hysteretic component of wingbeat load (primarily, aerodynamic dissipation) must be the result either of thoracic damping, and possibly the action of additional muscles (the *b1* muscle [89], *etc.*) acting independently of the main flight muscles.

If, for a moment, we neglect thoracic damping and additional musculature, we can consider a simple initial attempt to match muscular and load requirement loops. We scale the muscular load (*i.e.*, increase the estimated muscle mass) until the net power (area) of the two loops matches. Figure 8 illustrates this process: we denote the net-power-matched muscular work loop, the 'nominal' muscular work loop. There is a problem with this simplified approach: the actual load profile of the nominal muscular work loop (*i.e.*, an ellipse) does not exactly match the profile of the load requirement loop (Fig. 8A-B). For both synthetic and biological wing kinematics, there are stroke angles at which the load requirement loop is larger than the load supplied by the musculature: points at which the power generation of the muscle is not sufficient to generate the wingbeat motion. Despite the average power generation of the muscles being sufficient, their instantaneous load and power generation is insufficient.

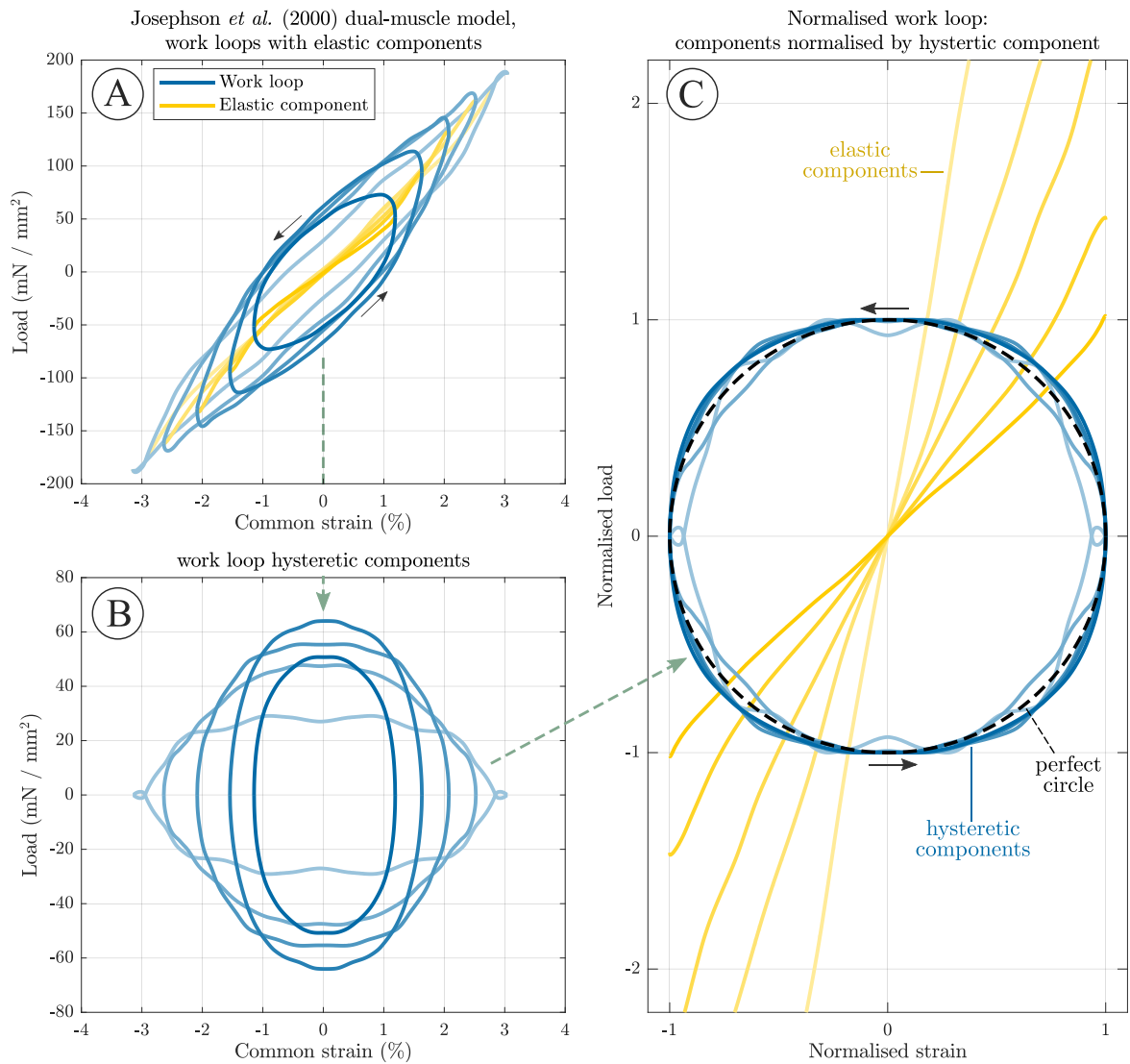

**Figure 7:** The elliptical nature of the dual-muscle model (§2.5). (**A**) Raw load-strain data, *i.e.* work loops. for the dual-muscle model, based on the empirical data of Josephson *et al.* [29]. (**B**) The hysteric component of the raw dual-muscle work loops. (**C**) Elastic and hysteretic components of the dual-muscle work loops, normalised by the hysteretic component in load, and illustrated against normalised strain. This normalisation illustrates the way in which all these these loops are accurately modelled by a circular normalised profile (elliptical unnormalised profile) irrespective of muscular strain amplitude.

To ensure that the flight motor instantaneous load and power generation is sufficient, larger primary flight muscles are required. In work-loop terms, only a muscular hysteretic work loop which is at least as great as the load requirement loop at every point in the stroke cycle will be sufficient in instantaneous power and load. Figure 8 shows these work loops – 'load-matched' loops – for the average wing load requirement profiles associated with synthetic (Fig. 8A) and biological (Fig. 8B) wingbeat kinematics[1]. Note that these load-matched loops now show areas where the instantaneous muscular load exceeds the load requirement. This excess load can be dissipated by additional dissipative effects in the system (*e.g.*, thoracic damping), whereas insufficient load cannot be remedied in the same way.

The load-matched muscular work loops in Fig. 8 are necessarily only approximations, but their implications are striking. In the case of the biological wingbeat kinematics, the load-matched work loop and the nominal work loop are quite similar: the load-matched mean loop shows only a 28% loss in power over the ideal net-work-matched state (Fig. 8B). Across the full dataset based on biological kinematics, this loss ranges from a best case of only 13%, for a datapoint of Shen *et al.* [24]; to a worst case of 80%, for a datapoint of Meng *et. al* [25]. That is, even the worst-case datapoint based on biological kinematics retains roughly twice as much useful work as the average for synthetic kinematics (20% vs 10%); and, as a whole, the biological kinematics perform several times better. Physically, wing aerodynamic dissipation under biological kinematics follows near-elliptical profiles, which are good matches to the elliptical muscular loop. In contrast, in the case of the synthetic wingbeat kinematics, load-matching is practically impossible: to match the instantaneous load and power requirements associated with strong vortex-capture drag peaks around the wing stroke extrema, the musculature requires a large excess. In such a situation, the muscular instantaneous load and power generation is too large over the vast majority of the stroke cycle, and so nearly 90% of the muscular power is wasted (dissipated). Even given the uncertainty in the aerodynamic data the core flight musculature is fundamentally unsuited to generating synthetic wingbeat kinematics.

We pause for a moment on this stark distinction between the behaviour of synthetic and biological wingbeat kinematics. This distinction stands in significant contrast to existing studies, which do not discern any significant functional difference, *e.g.*, in overall lift or power consumption [64,85,90], between biological and synthetic kinematics. This analysis of the drag load associated with wingbeat kinematics, and their relationship to muscular loads, provides another avenue towards understanding biological wingbeat kinematics. Here, we have evidence that biological kinematics are significantly better suited to the flight motor musculature: their load requirements form a near-elliptical loop, matching the musculature's elliptical work loop profile. Our data-synthesis analysis means that we cannot directly study the effect of specific wingbeat kinematic variables on these relationships, but we hypothesise that the wing elevation angle is a key factor in generating this elliptical loop. The wing elevation angle may alter the nature of the wing vortex-capture event, so that the peak drag load has different amplitude and timing (Fig. 8). If this is so, then the need to ensure flight motor load matching could explain the wing elevation angle variation in insect flight – further research is required.

*Overleaf:* **Figure 8:** Matching muscular load output and wing load requirements for two sets of wing kinematics: synthetic (**A**), and biological (**B**). Results are shown only for mean profiles, for clarity. Two approaches to matching muscular and wing loads are available: one based on matching nominal net power, and one based on matching instantaneous load and power. Matching nominal net power yields similar results for both sets of kinematics; but matching load yields radically different results, due to the non-elliptical load requirement loop under synthetic kinematics. This non-elliptical loop illustrates an important distinction between synthetic and biological wing kinematics: the *D. melanogaster* DVM and DLM are poorly suited to generating synthetic wingbeat kinematics, but well suited to generating observed biological wingbeat kinematics. This suggests that aspects of biological wing kinematics are specifically adapted to, or connected with, the dynamics of muscular actuation.

---

[1] As a practical feature, we neglect loads below a certain threshold (35% of each profile's peak load), to permit slight load mismatches at the junction between upstroke and downstroke. Attempting to match the exact change in load sign at this junction is futile, given uncertainty in the data and model.

## A  *Synthetic wingbeat kinematics*

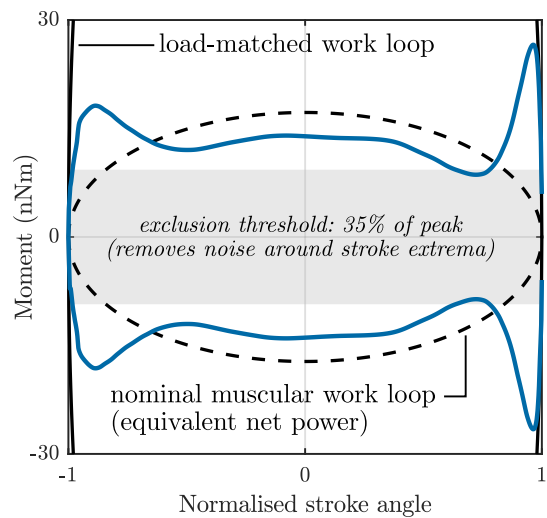
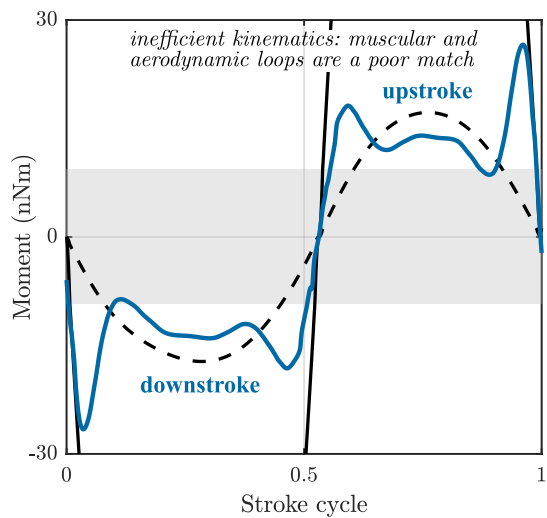
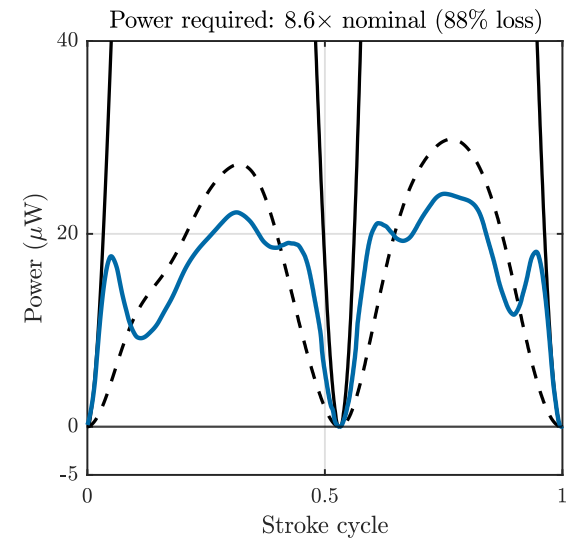

## B  *Biological wingbeat kinematics*

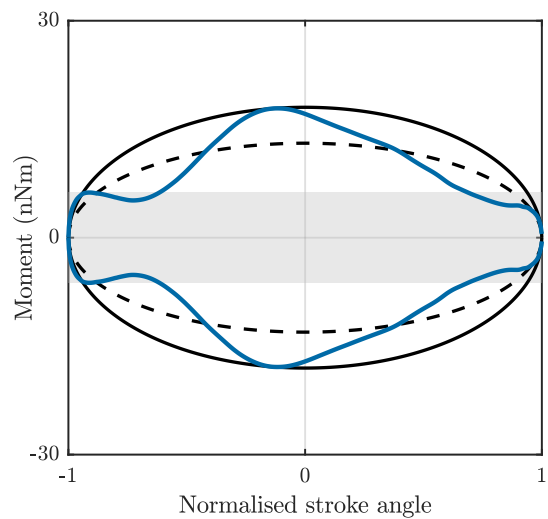
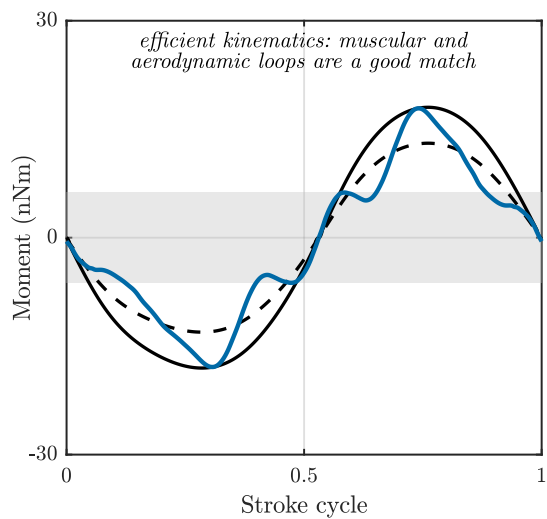
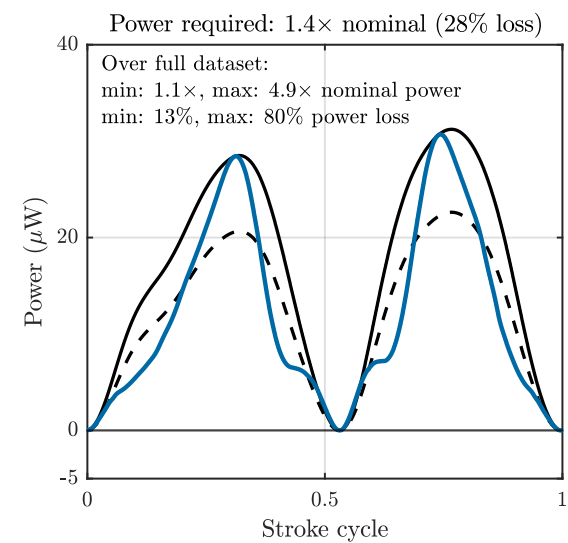

### 4.3. Intrinsic muscular elasticity provides the majority of flight motor elasticity

Having matched the hysteretic component of the muscular load and wing load requirement work loops (Fig. 8), we can turn to a complete match of hysteretic and elastic components. The overall scale factor of the muscular work loop is determined by the match between hysteretic components (net work and/or instantaneous power), and thus, the scale of the intrinsic muscular elasticity. We may then compare intrinsic muscular elasticity to the wing load requirement elastic component, and, potentially, attribute differences in these profiles to the modulation of specifically thoracic elasticity (§2.6-§2.7). In terms of wing load requirement, given the conclusions of §4.2 – that the musculature was largely incapable of generating the synthetic wingbeat kinematics, due to the stroke vortex capture drag peaks at the stroke extrema – we analyse only the biological wingbeat kinematics, and do so only for the dataset mean profile (*cf.* Fig. 8). In terms of muscular load generation, the estimates of Chan and Dickinson [9], that indicate that both DLM and DVM peak-to-peak strains lie within the range of 3.3-3.5%, though the strain waveform estimates associated with these results are coarse. To hedge the uncertainty in these strain estimates, we perform the match for two strain values within our source dataset, that of Josephson [29]: peak-to-peak amplitude 2.4%, and peak-to-peak amplitude 3.3%.

This yields four similar muscular work loops, matched to the single wing load requirement loop (Fig. 9). These matched loops show several striking features. Most significantly, the intrinsic elasticity of the muscles matches the elastic component of the wing load requirement loop well: there is no need for significant thoracic elasticity. Indeed, the intrinsic muscular elasticity tends to exceed the load requirement elasticity – the flight motor has stronger elasticity than required. If thoracic elasticity plays a role, it is likely to be related to small differences in profile; and not bulk energy absorption. However, the tendency toward excess muscular elasticity may indicate that inertial effects of the thorax (the inertia of muscles and organs) are more important than its elastic effects. Without further data we cannot quantify these thoracic effects in more detail, but the possibility of thoracic inertia playing a significant role in the resonant dynamics of *D. melanogaster* raises important methodological questions. Most modelling approaches to insect flight motor resonance consider the wings as the primary inertial load, and neglect the inertia of the thorax [42,91]. If thoracic inertia is significant in some of these insects, then resonant parameters – for instance, estimated resonant quality factors – may be inaccurate. Further experimental results could shed significant light on this topic. However, despite this uncertainty, our estimates broadly exclude any significant thoracic elasticity: muscular elasticity significant is at least strong enough – and possibly, too strong – to match for wing inertial effects.

*Overleaf:* **Figure 9:** Matching muscular load generation to wing load requirements. Four different matched muscular work loops can be identified: using nominal work matching, and load matching; and using muscular data at peak-to-peak strain amplitudes of 2.4% ('2%') and 3.3% ('3%'); in combination. For all of these matched loops, the intrinsic muscular elasticity is at least as great at the elasticity requirement. Thoracic elastic modulation is not required to ensure *D. melanogaster* flight motor consistency and/or resonance, except perhaps to resolve small differences in nonlinear elastic profile.

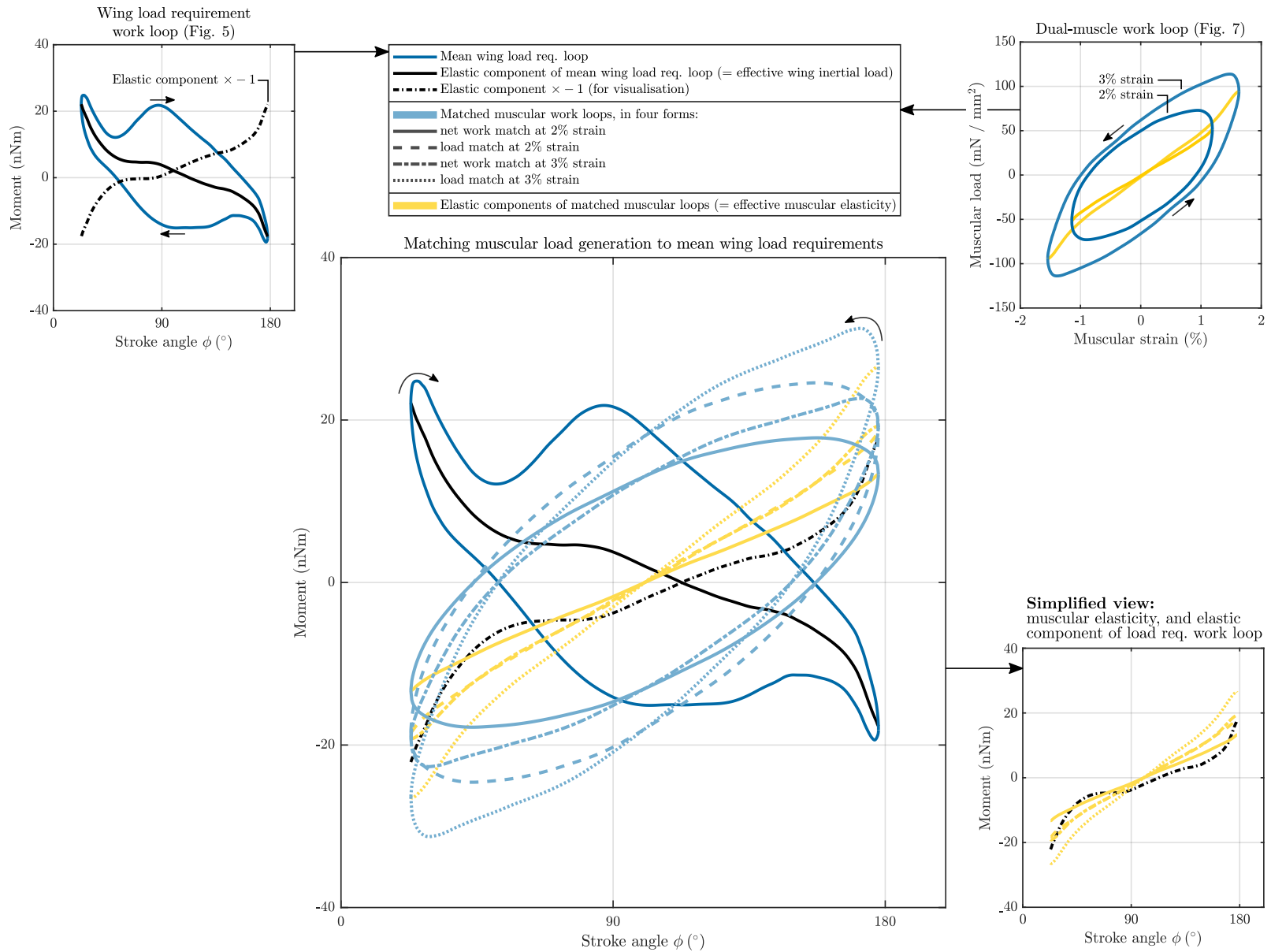

# 5. Discussion
## 5.1. Biological implications

The inverse-problem analysis set out in this work leads to several conclusions regarding the nature of thoracic elasticity in *D. melanogaster*. Firstly, we established that the energetic effect of overall flight motor elasticity (muscular and thoracic) in the flight motor of *D. melanogaster* can be significant, but is quite sensitive to the details of the wingbeat kinematics and aerodynamics (Fig. 5). Typical energy savings associated with flight motor resonance in the range of 0-30%, with averages on the order of 10%. The significance of this variation suggests that there might be an imperative to choose find wingbeat kinematics which assist (rather than, confound) the absorption of inertial effects by elasticity. We established that achieving this energetically-optimal state typically required nonlinear flight motor elasticity – a nonlinear elasticity which was, in fact, fundamentally related to the triangle-wave resemblance shown by *D. melanogaster* wingbeat kinematics (Fig. 4). These kinematic effects are likely to generalise: wingbeat kinematics in other insects are often not simple-harmonic [92,93], and so energetically-optimal flight motor elasticities in these insects are also likely to be nonlinear.

Secondly, in addition to its energetic effect, the effect of flight motor elasticity on the *distribution* of load requirement / generation within the wingbeat cycle was also found to be significant. Tuning this elasticity allowed significant alterations to the timing of peak load requirement and the symmetry of DVM and DLM actuation – all while maintaining overall energetic optimality (Fig 6). The role of elasticity (and, subtle nonlinearities in elasticity) in altering wing load timings, independent of any energetic effects, is a role which has not previously been recognised in the context of insect flight motor operation.

Indeed, thirdly, when comparing *D. melanogaster* wingbeat motion to muscular action, we found that the temporal distribution of load requirement within the wingbeat cycle had a significant effect on the feasibility of the flight motor operation. Biological wingbeat kinematics were well-suited to the flight motor musculature: the broadly elliptical work loops required matched well the dynamics of the musculature. Synthetic kinematics were not well-suited: strong vortex-capture drag peaks necessitated significantly larger muscles (~8× nominal) in order to generate sufficient instantaneous load and power, and wasted a significant proportion of their overall power output doing so (Fig. 8).

And finally, by matching wingbeat motion to muscular action, we found that the intrinsic elasticity of the *D. melanogaster* flight muscles was roughly equivalent to, and even somewhat in excess of, the elasticity required for consistency with the wing load requirement: there is no need for significant elastic modulation arising from the thoracic structure (Fig. 9). This challenges the supposition that specifically thoracic elasticity plays a key energetic role in the flight motor of *D. melanogaster*. It suggests instead that, if thoracic elasticity plays a role, it is likely to be related to small nonlinear effects rather than bulk energy absorption; and that the inertial effects of the thorax (the inertia of muscles and organs) may be more important than its elastic effects. This model of the role of the thoracic structure within the flight motor of *D. melanogaster* contrasts with other extant models that emphasis its role as an elastic modulator [4,46]. Indeed, this result lends further weight to earlier hypotheses of Alexander [94], Alexander and Bennet-Clark [95], and Ellington [96], that intrinsic muscular elasticity is a dominant elastic force within the flight motor of smaller insects, such as *D. melanogaster*.

Interestingly, there are several reasons to believe that this conclusion regarding negligible thoracic elasticity does not generalise to other insect species. According to existing semi-empirical estimates, fruit fly species show low resonant quality factors (*i.e.*, low levels of elasticity vis-à-vis dissipation) [43]. Insects with higher quality factors require stiffer elasticity relative to their net power output, and in these cases, muscular elasticity alone is unlikely to be sufficient. This scaling estimate is seemingly confirmed by existing experimental results for hawkmoths (*M. sexta*), which indicate that thoracic elasticity is significant w.r.t. muscular elasticity in these insects [44]. Interestingly, this leaves *D. melanogaster* – and potentially, similarly-scaled insects – in a distinct class. We propose that *D. melanogaster* lies on the boundary between insects with significant thoracic elasticity (*M. sexta*), and even smaller-scale insects (*e.g.*, *Paratuposa placentis*) where inertial effects themselves are practically insignificant [97]. For *D. melanogaster*, inertial effects are significant, but are largely accounted for by muscular, rather than thoracic elasticity. Indeed, the role of thoracic elasticity in *D. melanogaster*, if any, is much more likely to relate to load matching than resonant energy savings (*cf.* §3.2). This represents a practical manifestation of a fundamental principle in nonlinear dynamics: even in systems which are totally dissipation-dominated, energetically-neutral nonlinear elasticities can exist: elasticities which can serve to modulate the required input load waveform, at no energetic cost [74].

We note that there is uncertainty in our quantitative estimates of flight motor, muscular, and thoracic elasticity. Not only the uncertainty arising from within the source dataset (aerodynamic, kinematic, *etc.*) but also: we generalised asynchronous muscle behaviour across species (from *C. mutabilis* to *D. Melanogaster*); we assumed that empirical muscular load-vs-strain profiles would generalise across different muscular lengths undergoing equivalent strain profiles; and we assumed exact anti-symmetry in terms of DVM and DLM behaviour (identical mass and strain amplitude, phase offset by 180°). Together, these assumptions suggest approaches for constructing a more precise picture of the dynamics of the flight motors of *D. melanogaster* and other insect species. We propose an integrated study of wing kinematics (through high-speed cameras); aerodynamics (though CFD or similitude modelling); in-flight muscular strain estimates (through X-ray diffraction); and *ex vivo* muscular load estimates (through load-strain analysis). Developing techniques such as finite-element analysis of tomographic data for insect thoraxes [98] show significant potential for integration within this methodology. An integrated study of this form would be able to probe the nature of flight motor elasticity in *D. melanogaster*, or any other insect, in much finer detail.

## 5.2. Methodological implications

Our inverse-problem approach allows a characterisation of the nature of thoracic elasticity, and thoracic resonance, in insects, using a range of easily-observable data. In a broader biological context, this inverse-problem methodology shows potential for application to other small-scale biological structures, for which direct experimentation is difficult. In the specific context of insect flight, a key methodological comparison is between this inverse-problem approach, and a direct observation approach, in which experimental dynamic mechanical analysis (DMA) is used to characterise dynamic properties of the thorax directly. These two methodologies are complementary. An inverse-problem approach infers the elasticity directly experienced by the drivetrain, but relies on several modelling assumptions; experimental DMA observes the elasticity associated with a mode of thoracic deformation, but this mode of deformation may only partly represent the mode of thoracic deformation during flight. An interface between these two methods would represent a powerful dual technique for probing insect flight motor elasticity and resonance. In the case of *D. Melanogaster*, no DMA data is available – and such data is likely to be challenging to obtain given the small size of the insect – but data is available for other species, *e.g.*, *M. sexta* [44]. An extension to these insects is a key future application for the inverse-problem methodology.

As noted, in the course of our analysis, a few modelling assumptions were made. We assumed the flight motor to undergo parallel-elastic actuation [42,43], with a linear phase-matched transmission between muscular and wingbeat motion. Further data, *e.g.*, on phase differences between the musculature and wing [43], would allow these assumptions to be confirmed or further refined. Overall, the success of our inverse-problem characterisation of thoracic elasticity is founded on the extensive nature of existing literature datasets (*e.g.*, CFD studies, X-ray diffraction studies, ex vivo muscular studies, *etc.*). This success bodes well for the application of inverse-problem techniques to other insect species, *e.g.*, *M. sexta*, for which analogous datasets are also available. It also highlights the importance of these individual datasets to the interpretation of overall flight motor behaviour, and provides renewed motivation for the generation of further data – particularly, in the context of integrated studies of particular species.

### 5.3. Technological implications

Our inverse-problem approach led us to a characterisation of the nature and effect of flight motor elasticity in *D. Melanogaster*. This characterisation has implications both for the functional morphology of *D. Melanogaster*, and for the design of flapping-wing micro-air-vehicles (FW-MAVs). Several key technological implications can be noted. Firstly, energetically-optimal elasticities for flight motor systems can be nonlinear – and indeed, this nonlinearity is closely related to the waveform of the wing stroke oscillation. In FW-MAVs, nonlinear elasticity may lead to similar energetic benefits. Secondly, the role of elasticity in flight motor systems may be more than simply energetic: elasticity has a powerful role in controlling the load requirement waveform, and the timing of peak load requirement. For actuator well-suited to intermittent (*i.e.*, low duty-cycle) operation, nonlinear flight motor elasticity can favourably alter these load timings, even when energy savings associated with elasticity are small. These results suggest new design strategies to achieving efficient FW-MAVs: the use of tuned elasticities to generate the most efficient match between actuator load profiles (and timings of peak load generation) and wing load requirement profiles. And finally, the contribution of subtle features in flapping-wing kinematics to the wing load requirement waveform, and the significance of this waveform, have not previously been recognised. Our results illustrate how apparently innocuous wingbeat kinematics features – for instance, the wing elevation angle, which does not significantly alter mean lift or drag [62,64] – can be of key significance in ensuring that wing load requirement timings match an actuator's optimal load generation timings. As illustrated in §4.2, the energetic consequences of a poor timing match can be severe. These results suggest new approaches to optimising of FW-MAV wingbeat kinematics: and optimisation based on load waveform, and not only overall load.

## 4. Conclusions

In this work, we applied an inverse-problem approach to identify elastic properties of the flight motor of *D. melanogaster*. Synthesising a range of CFD, muscular-dynamic, and wingbeat-kinematic data, we were able to identify several novel properties of the *D. melanogaster* flight motor. We established that the energetic effect of flight motor elasticity in the flight motor of *D. melanogaster* can be significant, but that this effect is sensitive to the wingbeat kinematics and aerodynamics: elastic energy savings may range from 0-30%. Achieving maximal elastic energy savings required nonlinear elasticity – a nonlinear elasticity which is fundamentally related to the triangle-wave resemblance shown by *D. melanogaster* wingbeat kinematics. In addition, we demonstrated how careful tuning of this nonlinear elasticity allows significant control over the timing of peak load and power requirement, all while maintaining maximal energy savings. The role of elasticity in governing wing load timings is a role which has not previously been recognised in the context of insect flight motor operation.

And finally, when comparing *D. melanogaster* wingbeat motion to muscular action, we established two important features of overall flight motor operation. We established that, in terms of load generation, biological wingbeat kinematics were well-suited to the flight motor musculature, whereas synthetic kinematics were not – to the extent that the musculature is largely incapable of generating synthetic kinematics. This highlights again the energetic significance of apparently innocuous features in observed insect wingbeat kinematics – for instance, the variation in wing elevation angle. And we identified that the intrinsic elasticity of the *D. melanogaster* flight muscles was roughly equivalent to, and even somewhat in excess of, the elasticity required for consistency with the wing load requirement: there is no need for significant thoracic elastic modulation. Together, these newly-identified properties lead to novel conceptual model of the *D. melanogaster* flight motor: a strongly-nonlinear structure, highly adapted to the dynamics, load timings, and elasticity of the primary flight muscles. Our inverse-problem methodology sheds new light on the complex behavior of these tiny flight motors, and provides avenues for further studies in a range of other complex biological structures.